\documentclass[final,authoryear, times, 3p]{elsarticle}

\usepackage{lineno,hyperref}
\modulolinenumbers[5]

\makeatletter
\journal{Journal of Financial Stability}
\def\ps@pprintTitle{%
	\let\@oddhead\@empty
	\let\@evenhead\@empty
	\def\@oddfoot{}%
	\let\@evenfoot\@oddfoot}
\makeatother

\usepackage{numcompress}\biboptions{authoryear}

\usepackage{subcaption}
\usepackage{epstopdf}
\usepackage{graphicx} 
\usepackage{amsfonts}
\usepackage{amsmath} 
\usepackage{bm}
\usepackage{amssymb} 
\usepackage{hyperref} 
\usepackage{siunitx} 
\usepackage{booktabs} 
\usepackage{cleveref}
\usepackage{float}
\usepackage{multicol}
\usepackage{multirow}
\usepackage{booktabs}
\usepackage{tabu}
\usepackage{color}
\usepackage{todonotes}
\usepackage{lmodern}
\usepackage{siunitx}
\usepackage{soul}

\usepackage{natbib}
\defcitealias{baselcreditriskaccounting}{BCBS,~\citeyearpar{baselcreditriskaccounting}}

\defcitealias{guidencenpl}{BCBS,~\citeyearpar{guidencenpl}}

\defcitealias{basel1}{BCBS,~\citeyearpar{basel1}}
\defcitealias{basel2}{BCBS,~\citeyearpar{basel2}}
\defcitealias{basel3}{BCBS,~\citeyearpar{basel3}}

\defcitealias{baselRB}{BCBS,~\citeyearpar{baselRB}}
\defcitealias{baselClimateMeth}{BCBS,~\citeyearpar{baselClimateMeth}}

\defcitealias{baselClimatePrinciples}{BCBS,~\citeyearpar{baselClimatePrinciples}}

\defcitealias{baselClimateRiskTransmission}{BCBS,~\citeyearpar{baselClimateRiskTransmission}}

\usepackage[utf8]{inputenc}
\usepackage[T1]{fontenc}

\usepackage{mathtools}

\newdefinition{defi}{Definition}

\usepackage{color} 

\begin{document}

\begin{frontmatter}
\title{Estimating the impact of supply chain network contagion on financial stability}

\author[csh]{Zlata Tabachová} \ead{tabachova@csh.ac.at}
\author[csh,wu]{Christian Diem} \ead{diem@csh.ac.at} 
\author[cb,csh]{András Borsos} 
\author[cbd]{Csaba Burger} 
\author[csh,cosy,sfi]{Stefan Thurner\corref{cor}} \ead{stefan.thurner@meduniwien.ac.at}
\cortext[cor]{Corresponding author}

\address[csh]{Complexity Science Hub Vienna, Josefst\"adter Stra\ss e 39, A-1080, Austria}
\address[wu]{Institute for Finance, Banking and Insurance, WU Vienna University of Economics and Business, Welthandelsplatz 1, A-1020, Austria}
\address[cb]{Financial Systems Analysis, Central Bank of Hungary, Szabadság tér 9, Budapest 1054, Hungary} 
\address[cosy]{Section for Science of Complex Systems, Medical University of Vienna, Spitalgasse 23, A-1090, Austria} 
\address[sfi]{Santa Fe Institute, 1399 Hyde Park Road, Santa Fe, NM 87501, USA} 
\address[cbd]{Directorate Statistics, Central Bank of Hungary, Szabadság tér 9, Budapest 1054, Hungary} 
\address[ceu]{Department of Network and Data Science, Central European University, Quellenstrasse 51, 1100 Vienna Austria} 
 
\begin{abstract}
Realistic credit risk assessment, the estimation of losses from counterparty's failure, is central for the financial stability. Credit risk models focus on the financial conditions of borrowers and only marginally consider other risks from the real economy, supply chains in particular. Recent pandemics, geopolitical instabilities, and natural disasters demonstrated that supply chain shocks do contribute to large financial losses. Based on a unique nation-wide micro-dataset, containing practically all supply chain relations of all Hungarian firms, together with their bank loans, we estimate how firm-failures affect the supply chain network, leading to potentially additional firm defaults and additional financial losses. Within a multi-layer network framework we define a financial systemic risk index (FSRI) for every firm, quantifying these expected financial losses caused by its own- and all the secondary defaulting loans caused by supply chain network (SCN) shock propagation. We find a small fraction of firms carrying substantial financial systemic risk, affecting up to $16\%$ of the banking system's overall equity. These losses are predominantly caused by SCN contagion. For every bank we calculate the expected loss (EL), value at risk (VaR) and expected shortfall (ES), with and without accounting for SCN contagion. We find that SCN contagion amplifies the EL, VaR, and ES by a factor of $4.3$, $4.5$, and $3.2$, respectively. These findings indicate that for a more complete picture of financial stability and realistic credit risk assessment, SCN contagion needs to be considered. This newly quantified contagion channel is of potential relevance for regulators' future systemic risk assessments.
\end{abstract}


\end{frontmatter}

\section{Introduction} \label{introduction}

Credit risk (CR) assessment is central for banks to operate in an economically sustainable way. CR materialises when a counterparty is unable  or unwilling to fulfil its contractual debt obligations. In 1988 the \textit{Basel Committee on Banking Supervision} (BCBS) published the first \textit{Basel Accord} (\citetalias{basel1})  providing a set of minimum capital requirements that must be met by banks to guard against CR. These requirements have been refined also in the following accords \textit{Basel II} (2004) (\citetalias{basel2}) and \textit{Basel III} (2010) (\citetalias{basel3}), reflecting the importance of CR for financial stability. The regulatory path has been evolving towards a system where capital requirements are explicitly linked to the risk of activities undertaken by banks. This means that as perceived default risks increase, so do the risk-weights and the bank's capital buffer. Therefore, for being able to properly determine adequate capital buffers and adjust their interest rates, it is essential for banks to have an in-depth knowledge of all sources of CR.

Recent crises like the COVID-19 pandemic, the conflict in Ukraine, and several natural disasters have impressively shown that the propagation of shocks along supply chain networks (SCNs) lead to large financial losses of firms, and as a consequence, also to their creditors. It was shown that shock propagation along supply chains can be dramatic, for example in the case of Hurricane ``Katrina'' \citep{hallegatte2008adaptive}, the Japanese earthquake of 2011 \citep{carvalho2021supply, inoue2019firm}, natural disasters in the US\citep{barrot2016input}, the UK lockdowns during the COVID-19 crises \citep{pichler2022forecasting}, disruptions of natural gas supply \citep{brief2022austria}, or failures of systemically important firms \citep{diem2022quantifying}. The COVID-19 crises led to substantial shock propagation effects along SCNs and affected almost every sector, including the especially vulnerable small and medium enterprises (SMEs) \citep{bartik2020impact}. Banks only did not suffer extensive losses from non-performing loans during the pandemic by the virtue of enormous liquidity support measures, furlough schemes, and other measures imposed by governments and central banks \citep{milstein2021did, ECB}. In the future, an accelerating climate crisis is likely to increase the frequency and magnitude of natural disasters, which will make the propagation of shocks along supply chains an even larger concern \citep{willner2018global,battiston2021climate}.

To date, shock propagation through firm-level SCNs has not yet reached its ways into quantitative financial risk assessment of banks, or stress testing methodologies used by regulators; it has only recently been scarcely featured by research. Traditionally, credit risk models focus on the financial conditions of the borrowers, using their financial statements as the most relevant inputs. A rating assessment of borrowers usually involves variables derived from \textit{balance sheets}, \textit{income statements}, and \textit{cash flow statements}. For example, according to the BCBS, the \textit{leverage ratio} (\textit{total assets} divided by \textit{total equity}) has the strongest explanatory power, especially when combined with \textit{revenue}, see \citetalias{baselRB}. Credit rating models that are applied on these variables, are primarily based on classical statistical methods, such as \textit{logistic regression} \cite{cox1958regression}\footnote{By now, machine learning and deep learning based credit risk models outperform classical ones in terms of accuracy \citep{shi2022machine}. They can handle large alternative data sets, such as text- or graph data \cite{hamilton2020graph} including the extraction of relevant information from these. These approaches could be used to include SCN data. As discussed in the \textit{European Banking Authority report on big data and advanced analytics} \cite{ebaBDAA}, there is an evident tendency in employing  \textit{big data and advanced analytics (BD$\&$AA)} into many aspects of the banking business, such as fraud detection or client interactions. There is also growing interest in the area of financial  risk management. On the other side, there are still many concerns like the ``biasness'' or the fact that (BD$\&$AA) methods still largely remain ``black boxes''. That is why regulators so far have been cautious of approving their usage for risk management.}. Credit risk management evaluates information about suppliers and buyers in given markets or under specific economic conditions in, predominantly, qualitative ways \citep{gorgijevska2019qualitative, moretto2019supply}. The subjective analysis of clients' supply chain exposures can complement existing quantitative credit ratings, but is limited to first-tier supplier-buyer relations and, hence, can not capture the complex structures of SCNs that transmit disruptions far beyond the first tier \citep{carvalho2021supply}. 

Research on network-based financial systemic risk \citep{boss2004network,battiston2012debtrank, thurner2013debtrank,  poledna2015the, poledna2018quantification, diem2020minimal, feinstein2017measures, gai2010contagion,  thurner2022csp, poledna2018identifying}, and macro prudential stress testing \citep{farmer2022handbook, cont2017fire, glasserman2016contagion, gauthier2012macroprudential, cont2010network, elsinger2006risk}, primarily focuses on the nature of the propagation of shocks in \textit{financial} networks, but does not include contagion effects along the SCNs, the back bone of the real economy. Further examples of this stream of literature include \cite{buncic2013macroprudential}, \cite{acharya2014testing}, \cite{levy2015dynamical}, \cite{ARNOLD20123125}, \cite{borio2014stress}, \cite{vazquez2012macro}. As pointed out in \cite{battiston2018financial}, the literature connecting the real economy with the financial system is strongly under-researched. \cite{herring2022objectives} and \cite{potter2020stressing} emphasise that the framework of current stress testing should be broadened to include non-financial threats to financial stability, such as economic or climate shocks. Similarly,  \cite{farmer2021stress}  highlights the need for so-called \textit{system-wide stress testing} to account for relations between economic and financial systems. An earlier attempt of agent based approaches to linking the real economy to the financial system is found in \cite{klimek2015bail}. Several regulating bodies and institutions such as the BCBS \citetalias{baselClimateMeth}, \citetalias{baselClimateRiskTransmission}, \citetalias{baselClimatePrinciples}, the Federal Reserve (FED) \citep{brunetti2021climate}, and the European Central Bank (ECB) \citep{ECBcreditrisk} pointed out the significance of climate-related risk to financial stability and identified supply chains as one of the relevant risk transmission channels. 

Due to a lack of granular data, so far it has been impossible to quantify exposures of financial systems to contagion in SCNs on the firm-level. On the industry level contributions in this direction were provided by \cite{guth2020modeling} where aggregated supply chain networks in the form of IO tables were used to assess the impact of supply chain disruptions in the context of the COVID-19 crises on the Austrian banking system. However, the use of aggregated industry-level production network data can cause substantial mis-estimations of production losses \citep{diem2023estimating}, showing the need for firm-level modelling approaches of supply chain shock propagation. Recently, the spreading of liquidity shortages on the production network \citep{huremovic2020production, Demir} was explored, which can also generate feedback effects for the financial sector \citep{silva2018bank}. The first works considering explicit interactions between the firm-level production network and the financial sector using granular network data for both systems are \citep{huremovic2020production, borsos2020shock}, however, the framework is limited to shocks that originate in the financial sector.

Here we present a data-driven computational framework for estimating how the initial failure of firms spreads along the supply chain network (SCN) within a country, leading to potentially additional firm defaults. These spread to the financial system through additional firm-loan write-offs and equity losses for banks. Based on a nation-wide micro-dataset, containing all supply chain links of all Hungarian firms in combination with  comprehensive credit registry data containing all commercial loans that firms obtained from banks, we are in the unique position to address the following two research questions. First, to what extent does systemic risk in the real economy --- created by cascades of production failures in national SCNs --- affect financial stability? And second, how is the credit risk exposure of banks --- measured by expected loss, value at risk, and expected shortfall --- amplified by contagion in these SCNs, or, equivalently, how much is credit risk underestimated by ignoring supply chain contagion? To answer these questions, we take two perspectives, a system-wide and one that is bank-specific. For the former, we introduce for every firm a \textit{financial systemic risk index} (FSRI). FSRI quantifies the financial losses of the banking system, caused directly by the firm's failure and indirectly from the resulting propagation of shocks in the supply chain network. For the latter, we then stress the system with $10,000$ different initial shock scenarios (each shock is an iid draw from firms' probability of defaults) and estimate additional risks for every single bank by generating loss distributions that take SCN contagion into account. In particular, we compute the \textit{expected loss} (EL), the \textit{value at risk} (VaR), and the \textit{expected shortfall} (ES) for all banks with and without SCN contagion.
In this way we extend the economic systemic risk index (ESRI) \citep{diem2022quantifying} --- that quantifies the supply-network-wide production (output) losses caused by the initial failure of a single firm or group of firms  --- to account for financial losses and potential contagion effects to the financial system.

\begin{figure}[t]
	\vspace{-1cm}
	\centering
	\includegraphics[scale=.4, keepaspectratio]{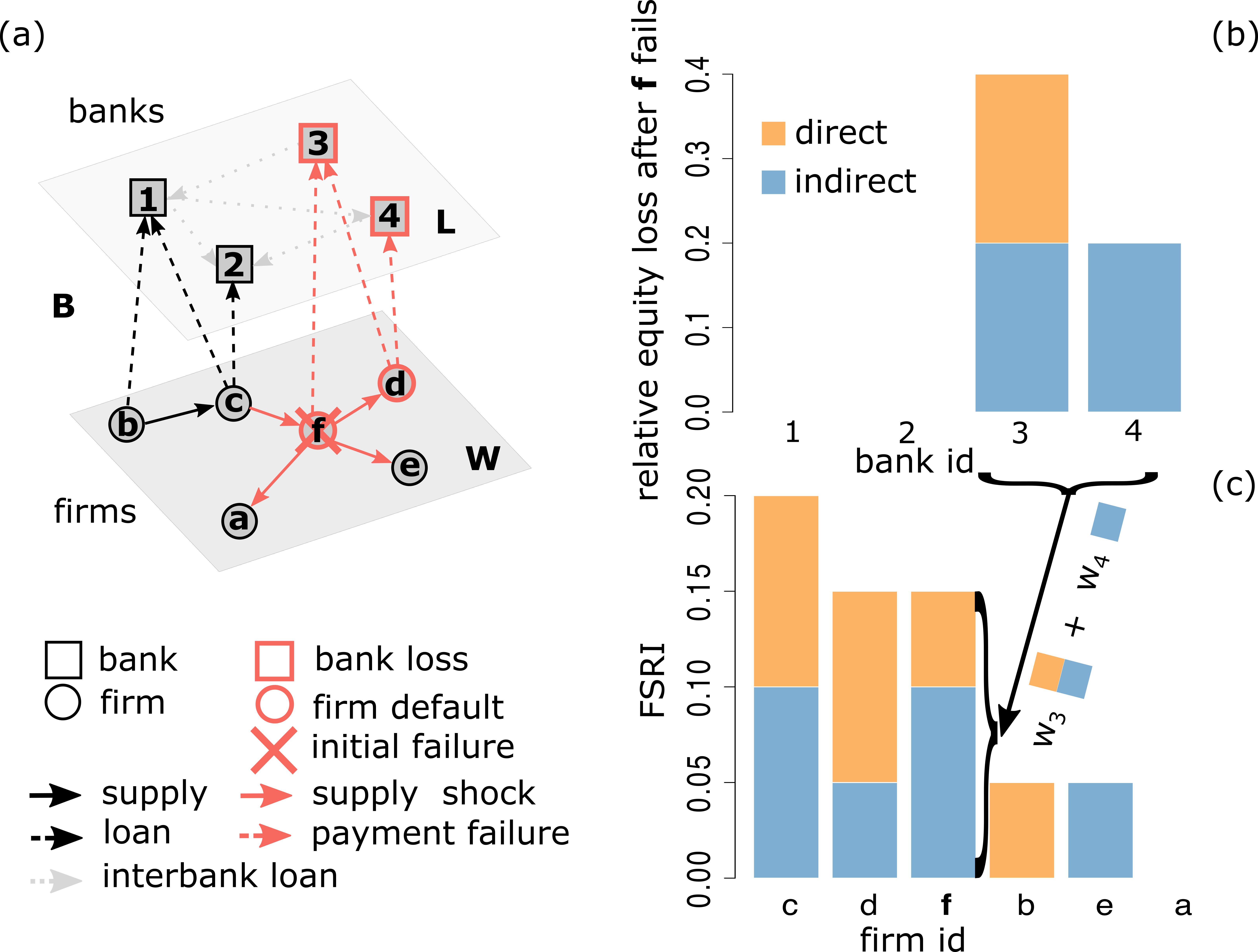}
	\caption{\textbf{Schematic view of contagion in the firm-bank multi-layer network.} \textbf{(a)} The bottom layer, $W$, represents  the supply chain network. Every node is a firm (circle) and a link, $W_{ij}$, (full arrow) represent a supplier-buyer relation, from supplier $i$ to buyer $j$. The top layer represents the inter-bank network, $L$. Nodes are banks (square) and a link, $L_{k\ell}$, (dotted arrow) is a liability bank $k$ has to bank $\ell$. Inter-bank links are not considered here, $L_{k\ell}=0$.  Layer $L$ is connected with the supply chain network, $W$, through bank-firm loans given by the matrix, $B$,  (dashed arrows). $B_{ik}$ is a liability of firm $i$ to bank $k$. In this example, we show how contagion starts with the initial failure of firm $f$ (red X) (assuming a $100\%$ production disruption), and spreads upstream to firm $c$ and downstream to firms $a$, $d$, and $e$ (full red arrows). As a result, production levels of those firms are reduced, leading to financial losses for firms firms $f$ and $d$, that cause their default (red circles). Whenever a firm defaults, the banks write off the corresponding loans in their balance sheets, resulting in losses of equity and (possibly) liquidity in the financial layer. \textbf{(b)} losses from the default of $f$ and $d$ are presented as bar-plot, with bank indices on the $x$-axis and the relative equity losses of the involved banks on the $y$-axis. In the case of the initial failure of $f$, bank 4 suffers a $20\%$ loss of equity indirectly (blue bar). Similarly, bank 3 suffers a $20\%$ loss of equity directly from the loan to firm $f$, and an additional $20\%$ loss from the additional default of firm $d$ --- we call this an indirect loss from supply chain contagion. \textbf{(c)} Overall losses in the bank layer are shown as a bar-plot  with firm indices on the $x$-axis and financial systemic risk index (FSRI) on the $y$-axis. FSRI is the weighted average equity loss of all banks, with the weights given by the relative equity size of each bank, i.e. $w_3=w_4=0.25$, since we assume that all banks in this example have the same equity; see main text. For the third bar we see, that FSRI of firm $f$ is equal to $15\%$, meaning that $15\%$ of total equity in the financial sector can be lost if $f$ fails. Analogically, firms $c$, $d$, $b$, $e$, and $a$ can cause $20\%,$ $15\%$, $5\%$, $5\%$, and $0\%$ of damage,  respectively. Financial contagion between banks is not considered. 
 }
	\label{twolayernetwork}       
\end{figure}

\section{Model and Methods} \label{method}

To describe how the propagation of production disruptions along supply chains spreads to banks, we use the supply chain network (SCN) of $n$ firms given by the $n \times n$ matrix, $W$. Its elements, $W_{ij}$, are the sales volumes of firm $i$ to firm $j$, or equivalently, purchases of firm $j$ from $i$ in monetary units (Euros per year).  The second network layer is the interbank network, $L$,  containing $m$ banks, where a link, $L_{ij}$, represents a liability of bank $i$ towards bank $j$. Note that $L_{ij}$ could also represent any other direct exposure, such as studied in \cite{poledna2015the}. Every bank $k$ is endowed with an equity, $e_k$. The two network layers are connected through the $n \times m$ firm-bank loan matrix, $B$, where $B_{ik}$ is the \textit{outstanding amount} of the liability firm $i$ has towards bank $k$. In this paper we focus on contagion from the supply chain, $W$, to the interbank network, $L$. We omit potential cascading effects inside $L$ \footnote{Interbank financial contagion channels could be straight forwardly included as an extension of the model presented in this work.}. This model is calibrated with actual data derived from Hungarian VAT and balance sheet data for the year 2019 for $W$, $L$, $B$, and $e$. For details, see Section \ref{data}. 

Schematically we describe the two-layer network in Fig.\ref{twolayernetwork}(a). The bottom layer represents the SCN, $W$, consisting of $6$ firms (circles). Arrows between them indicate supplier-buyer relations (from supplier to buyer), e.g. $b$ is a supplier of $c$ and $W_{bc}$ are the sales from $b$ to $c$ in Euros per year. The upper layer represents a financial network consisting of $4$ banks (squares). Layers are connected by firm-bank loans, $B$  (dashed arrows), e.g. firm $f$ has a loan from bank $3$, with $B_{f3}$ being the outstanding amount. All firm loans are assumed to be of size $B_{ik}=1$ and all interlayer links are, $L_{k\ell}=0$ (light gray; no contagion between banks is possible). The capital or equity, $e_k$ of all banks is set to 5.

Contagion spreading from the SCN layer, $W$, to the individual banks occurs in three steps: 
\begin{enumerate}
    \item quantify the propagation of  production disruptions in the SCN after a firm's initial failure;
    \item update equity and liquidity buffers of firms in response to potentially reduced production levels and check if they became insolvent and default on their loans;
    \item update the banks' equity buffers by writing off the loans of the initially failed (and defaulted) firms, and the firms that defaulted due to supply chain network contagion. 
\end{enumerate}
In this multi-layer network micro-simulation model, the actual data for $W$, $L$, $B$, and $e$ is considered to be the realisation of an ``undisrupted'' economy, where firms normally buy and sell in the SCN. The data contains  firms' income statements, balance sheets and the outstanding principles of banks' commercial loan portfolios of the respective year. Given this initial state of the economy, we can next simulate the effects on this economy when it is hit by an initial shock that affects the production capacities of firms and propagates through the supply chain network. This will now lead to counterfactual realisations of income statements and balance sheets of the firms. From these we infer which firms would have defaulted on their loans if the initial shock had happened. Banks write off the respective loans and suffer hypothetical losses, given the initial firm shock. We next describe the three steps of the model in more detail. 

\paragraph{Step 1} \label{step1} 
To simulate how a specific firm failure spreads through the SCN, $W$, we use the firm-level shock propagation mechanism of \cite{diem2022quantifying}, described in  \ref{appendixA}. There, a \textit{Generalized Leontief Production Function} (GLPF) is specifically calibrated  for every firm, $i$, to determine how much it can still produce if one of its suppliers, $j$, fails to deliver its products. We refer to this as a supply shock or downstream shock propagation (from supplier to buyer). If a customer, $\ell$, of firm $i$ stops buying it faces a demand shock or upstream shock propagation (from buyer to supplier). The undisturbed production state of the SCN at time, $t_0$, is represented by the production-level vector,  $h(t_0)$. Initially, each firm, $i$, produces $100\%$ of its original production level, i.e. $h_i(t_0) = 1$ for every $i\in \{1,2,\cdots,n\}$.  At time, $t_1$, an initial event occurs that disrupts the production of at least one firm in the SCN. This can be for example the (temporary) failure of a single firm,
or a simultaneous production stop of many firms due to a COVID-19 lockdown or a natural disaster. The vector $\psi \in \left[0,1\right]^n $ denotes the severity of the initial shock by specifying the remaining production level for every firm after the initial event occurred. For instance, $\psi_i=0.6$ means that firm $i$ can still produce $60\%$ of its pre-shock production level. We set the production level of every firm, $i$, to $h_i(t_1)=\psi_i$ ($h(t_1)=\psi$) \footnote{Note that initial shocks could affect firms at different time points, i.e. $\psi_i(t_{i})$ happens before $\psi_j(t_{j})$ if $t_{i} < t_{j}$. Here we always assume that $t_{i}=t_{j} = t_1 \; \forall \; ij$ and omit the time index in the notation. The model can be easily adjusted to feature different time points of disruption events. In the presented model the time intervals between two elements of the sequence $t_1, t_2, t_3, \dots, T$ correspond to  the same time unit, e.g., a calendar day or a calendar week.}. The initial shock, $\psi$, triggers an upstream and downstream shock propagation cascade along the supply chain links of firms. The propagation continues until every firm stops to adjust its production levels, i.e. $h_i(t_\tau) - h_i(t_{\tau +1}) < \epsilon$, and the system converges to a new stationary state at time $T := t_{\tau+1}$. This state is represented by the final remaining production levels, $h(T) \in \left[0,1\right]^n$, after the initial event. For example, $h_i(T)=0.3$ means that firm $i$ lost $70\%$ of its original production level.  We assume that production levels remain at this new state for a certain amount of time before firms can recover by finding new suppliers and buyers. Here, we do not model this rewiring of the network since a realistic rewiring dynamics needs substantially more assumptions and data that would go beyond the scope of this work. Therefore, the model outputs represent the short term effects before firms can adjust their production and sales structure.

In the example of Fig.\ref{twolayernetwork}(a) the initial shock scenario assumes the failure of firm $f$ (red X), e.g. due to an exogenous event such as a fire. Initially, we have $h(t_0)=(1,1,1,1,1,1)$ corresponding to a $100\%$ production level of all 6 firms $a$, $b$, $c$, $d$, $e$, and $f$, respectively. Next,  the initial shock at time $t_1$ is introduced with the vector,  $\psi=h(t_1)=(1,1,1,1,1,0)$, where the 6th element indicates the drop of $f$'s production level to $0\%$. The shock now spreads (red arrows) upstream to $f$'s supplier $c$ and downstream to its buyers $a$, $e$, and $d$. To keep the example short, we assume here that the production of $a$, $c$, and $e$ does not depend on the inputs of $f$, but $d$ can not produce without them, i.e. suffers a $100\%$ production disruption. Consequently, after convergence at, $T$, the final production level vector is $h(T)=(1,1,1,1,0,0)$. 

\paragraph{Step 2} \label{step2} 
The now reduced production levels, $h(T)$, cause a drop in firms' revenues and material costs that affects their profits. This change is relative to the pre-shock profits recorded in the income statement. Reduced profits  may turn into large losses and potentially will affect the equity and liquidity of firms, and eventually, the ability to repay their bank loans (if they have any). To find out whether a company can withstand the financial losses from a given production disruption we use the information from its income statement and balance sheet, a stylized version of which  is depicted in Table \ref{table:IncStBalSh}. 
\begin{table}[tb]
    \caption{Stylized balance sheet and income statement of firm, $i$.}
    \label{table:IncStBalSh}
    \begin{subtable}[h]{0.45\textwidth}
    \begin{center}
            \begin{tabular}{l|l}
            \textbf{Assets} & \textbf{Liabilities}\\       
            \hline
            \textrm{    }Short term assets $(a_i)$  &\textrm{    } Short term liab. $(s_i)$ \\ 
            \textrm{    }Inventory &\textrm{    } Other liabilities \\ \cline{2-2}
            \textrm{    }Fixed assets & \textbf{Equity} $(z_i)$ \\ \cline{2-2}
            \textrm{    }Other assets & \textrm{    }Profit $(p_i)$\\
             &  \textrm{    }Other equity\\
           \\[-1.9ex]\hline 
            \hline \\[-1.9ex]
            \textbf{Total Assets} & \textbf{Total Liabilities} \\
            \hline 
            \end{tabular}
    \end{center}
    \end{subtable}
    \hfill
    \begin{subtable}[h]{0.45\textwidth}
    \begin{center}
            \begin{tabular}{l}   

           \textbf{Operational profit}\\
            \textrm{        } + Revenues $(r_i)$ \\ 
            \textrm{        }  $-$  Material costs $(c_i)$\\    
            \textbf{Non-operational profit} ($o_i$)\\
             \textrm{        }  + Other income \\
             \textrm{        }  $-$ Other costs\\

            \\[-1.8ex]\hline 
            \hline \\[-1.8ex]
            \textbf{Net Profit $(p_i)$}\\
            \hline 
            \end{tabular}
        \end{center}
    \end{subtable}
\end{table}
For every firm, $i$, we infer for a given year its \textit{revenue}, $r_i$, its \textit{material costs}, $c_i$,  its \textit{non-operational profit}, $o_i$, consisting of \textit{other income} (e.g., income from financial assets, sales of land, etc.), and \textit{other costs} (e.g. HR costs, depreciation, rent, and other costs that can not be adjusted within a short period of time), from its income statement data. From the balance sheet data we read off firm $i$'s \textit{equity}, $z_i$, its \textit{short-term assets}, $a_i$ (including cash, short-term claims, and securities), and its \textit{short-term liabilities}, $s_i$ (including short-term loans to suppliers, etc.). 

We next determine how a  production reduction of firm, $i$,  translates into a reduction of its revenues, $r_i$, and its material costs, $c_i$,  affecting the profit, $p_i$. We assume that the non-operational profit stays the same, and that the revenue is reduced to $h_i(T) r_i$. Note, that for simplicity we neglect potential inventories of inputs and finished goods. We further assume that firms can reduce their material costs proportionally to the production level, i.e. material costs after the cascade are $h_i(T) c_i$. This assumes that firms only have to buy exactly the amount they need for production and neglects the fact that supply contracts might be long-term. This assumption might be overly optimistic. Since, the net profit is the sum of revenue, negative costs, and non-operational profit, $p_i = r_i - c_i + o_i$, the new profit at the reduced production level is $\tilde{p}_i = h_i(T) (r_i - c_i) + o_i$ and the change in profit is 
\begin{equation}
    \Delta p_i = p_i-\tilde{p}_i= \left( 1-h_i(T) \right)(r_i - c_i) \quad .
\end{equation}
The change in profits affects a firm's equity and its liquid assets (via the cash flow statement). The equity, $z_i$, of firm $i$ includes the equity from the previous year and the net profit of the current year, $p_i$. Note, that $p_i$ can also be negative. After accounting for the change in profits the new equity is $\tilde{z}_i = z_i - \Delta p_i$. According to The Insolvency Code \citep{Act} (Section 27 (2f))  firm $i$ becomes insolvent if $ \tilde{z}_i < 0$. The cash position of the firm at the balance sheet reporting date depends on the profits, $p_i$, made during this period\footnote{The explicit link between profits and cash is the cash flow statement. The indirect method for calculating the cash flow during the year starts with the profit $p_i$ and adds or subtracts transactions that are not cash effective and eventually yields the change in cash position. Examples for non cash effective transactions are adding back depreciation, adding a reduction of accounts receivable, deducting an increase in inventory levels, or deducting the purchase price of a new machine.}. A reduction of profits will cause a decrease of the firm's cash position. If  cash turns negative the firm becomes illiquid, can no longer pay bills or repay loans, and may become bankrupt according to the Insolvency Code \citep{Act} (Section 27 (2a)). We define the variable to measure a liquidity default more broadly as the sum of short-term assets (that could be sold fast) minus the sum of short-term liabilities (that are due soon), i.e. $a_i - s_i $. After the shock, firm $i$ has a liquidity of  $a_i - s_i - \Delta p_i$ and if it is negative we assume that it declares bankruptcy, defaults on its loan and banks write off all loans extended to $i$. Due to the cash flow calculation procedure and the assumption that the other transactions happen in the same way as in the case where the shock did not occur, the change in liquidity is $\Delta p_i$. This assumption presumably leads to an over-estimation of the liquidity loss as firms in distress would possibly defer investments into new machinery that are cash flow negative or could  try to receive additional loans from banks (cash flow positive) to avoid becoming illiquid. We abstain from modelling this behaviour in detail as it involves a number of additional assumptions that can not be backed with the available data. 

In summary, we derived two insolvency conditions for firms that depend on the size of the firm's production reductions, $1-h_i(T)$, and its financial strength (liquidity, $a_i-s_i$, and equity buffers, $z_i$). We assume that insolvency of a firm leads to its bankruptcy, which reflects the unlikeliness of loans to be repaid. The \textit{unlikeliness to pay indicator} is one of the conditions under which a bank classifies its client as defaulted, see, e.g., Article 178(1-3) of Regulation (EU) No 575/2013 \citep{regulation}. To conclude step 2, for all firms we define a binary default indicator 
\begin{equation}
  \chi_i (\psi) = \begin{cases} 
  1 \qquad  \text{if} \quad (z_i - \Delta p_i) \leq 0  \quad \text{or} \;\;  \quad  (a_i - s_i - \Delta p_i) \leq 0 \quad ,\\ 
  0 \qquad \text{if} \quad  (z_i - \Delta p_i) > 0  \quad \text{and} \quad   (a_i - s_i - \Delta p_i) > 0 \quad . \end{cases}
  \label{chi}
\end{equation}
It indicates, which firms defaulted ($\chi_i=1$) as a result of an initial shock, $\psi$, and the ensuing supply chain contagion after the cascade stops at time $T$. $\chi_i=0$ means $i$ is still going concern (not defaulted). To see the role of the equity- and liquidity default conditions in Eq. (\ref{chi}) individually, we define the equity default indicator vector, $\chi_{\rm{eq}}$,  that only considers the condition $ (z_i - \Delta p_i) \leq 0$, and the liquidity default indicator vector, $\chi_{\rm{l}}$ that only checks the condition $(a_i - s_i - \Delta p_i) \leq 0$. Note, that the union of the two vectors gives $\chi$. In the following we define all equations in terms of $\chi$, however, all definitions can be rewritten in terms of $\chi_{\textrm{eq}}$ and $\chi_{\textrm{l}}$ to analyse the effects of liquidity or equity defaults individually.

\paragraph{Step 3}\label{step3} 
Loans of defaulted firms are classified as \textit{non-performing loans} (NPL). We assume that borrowers are not subject to any forbearance solutions (as would be standard procedure, see \citetalias{guidencenpl}), and they cause write-offs or provisions for banks that reduce the banks' equity buffers. Impairment allowances for non-performing loans (or \textit{loan loss provisions}) implied by current accounting practice are described in the International Financial Reporting Standards (IFSR 9) \citep{IFRS9} or in \textit{Guidance on credit risk and accounting for expected credit losses} (\citetalias{baselcreditriskaccounting}). For simplicity, we assume a loss given default of 100\%, i.e., the entire exposure, $B_{ik}$, bank $k$ has towards firm $i$ at the time of default is written off. We distinguish between \textit{direct} and \textit{indirect} losses of banks' equity. \textit{Direct losses} are caused when a debtor, $i$, immediately defaults in response to the initial disruption, i.e., $0\leq \psi_i <1$ and $\chi_i=1$. 
Accordingly, we define the \textit{direct default indicator} via binary $n$-dimensional vector $\chi^{\rm dir} = \chi^{\rm dir}(\psi)$. \textit{Indirect} bank losses are caused when a firm $i$ defaults, $\chi_i=1$, but not due to its initial production disruption, i.e. $\psi_i=1$, but because of the additional production losses stemming from the propagation of the initial disruption. These equity losses of banks would not occur if supply chain network contagion is not present.  The \textit{indirect default indicator} $\chi^\text{indir}=\chi^\text{indir}(\psi)$ is defined as the binary $n$-dimensional vector
\begin{equation}
    \chi^{\textrm{indir}}_i= 
    \begin{cases} 
  1 \qquad  \text{if} \; \chi_i=1 \text{ and } \psi_i =1 \quad , \\ 
  0 \qquad \text{otherwise} \quad . \\ 
  \end{cases}
\end{equation} 
Then, $\chi^{\rm{dir}}_i =\chi_i -\chi^{\rm{indir}}_i$ holds\footnote{Note, that to obtain $\chi^{\text{dir}}$ we don't need to perform contagion at all, meaning that  \hyperref[step1]{Step 1} can be omitted, and in  \hyperref[step2]{Step 2} we put $h(T)=\psi$. $\chi^{\text{dir}}$ is obtained accordingly from Eq. (\ref{chi}).}. The fraction of lost equity, $\mathcal{L}_k$, of bank, $k$, --- after the initial shock,  $\psi$, has propagated through the supply chain network ---, is calculated as
\begin{equation}
    \mathcal{L}_k \equiv \sum_{j=1}^n \chi_j\frac{B_{jk}}{e_k}=\sum_{j=1}^n \left(\chi_j^{\rm{dir}}+\chi_j^{\rm{indir}}\right) \frac{B_{jk}}{e_k} \quad ,
    \label{lossesLambdak}
\end{equation}
where $\mathcal{L}_k=\mathcal{L}_k(\chi(\psi))$ is calculated for every bank $k\in \{1,2,...,m\}$. Since, the elements of $B$ are given in monetary units, we divide the loans, $B_{jk}$, by the respective bank's equity, $e_k$. The fraction of the lost equity from direct defaults is $ \mathcal{L}_k^{\rm{dir}} = \sum_{j=1}^n\chi_j^{\rm{dir}} \frac{B_{jk}}{e_k} $  and from the supply chain contagion induced defaults $\mathcal{L}_k^{\rm{indir}} = \sum_{j=1}^n \chi_j^{\rm{indir}}\frac{B_{jk}}{e_k} $. The equity losses suffered by the entire banking sector from initial shock, $\psi$, is 
\begin{equation}
    \mathcal{L}(\psi) \equiv \frac{\sum_{k=1}^m \mathcal{L}_k(\psi)e_k}{\sum_{\ell=1}^m e_{\ell}} \quad .
    \label{system_loss}
\end{equation}

To illustrate steps $2$ and $3$ in Fig.\ref{twolayernetwork}(a) the red circles indicate that firms $f$ and $d$ defaulted, i.e $\chi=(0,0,0,1,0,1)$. We consider that firms $a$, $b$, $c$, and $e$ were able to withstand the shock, $\psi=(1,1,1,1,1,0)$ caused by the initial disruption of $f$. Red dotted arrows indicate payment failures of defaulted firms $f$ and $d$, red squares show that banks $3$ and $4$ suffered losses. Bank $3$ suffered losses from two non-performing loans, $B_{f3}$ (directly) and $B_{d3}$ (indirectly). Bank $4$ suffered losses only indirectly from $B_{d4}$. For this scenario, $\chi^{\rm{indir}}(\psi)=(0,0,0,1,0,0)$ and $\chi^{\rm{dir}}(\psi)=(0,0,0,0,0,1)$, $\chi=\chi^{\rm{indir}}+\chi^{\rm{dir}}$. Recall, in this example all banks have the same initial amount of equity of $5$, and all loans have  the same outstanding amounts of $1$. Relative losses of banks after the initial failure of firm $f$ are depicted in Fig.\ref{twolayernetwork}(b), where every bar is associated to one bank. Banks $1$ and $2$ didn't experience any losses, explaining why $\mathcal{L}_1(\psi)=\mathcal{L}_2(\psi)=0$. Bank $3$ lost $1/5$ of equity directly (orange brick) and $1/5$ of its equity indirectly (blue brick), i.e. $\mathcal{L}_3(\psi)=\mathcal{L}_3^{\rm{dir}}+\mathcal{L}_3^{\rm{indir}}=2/5$. Similarly, we get $\mathcal{L}_4=\mathcal{L}_4^{\rm{indir}}(\psi)=0.2$ and $\mathcal{L}(\psi)=(3/20)=0.15$. 
\\

\textbf{ Estimating the effects of individual-firm failures on overall bank equity.} 
To address our first research question, to what extent a firm's systemic risk in the real economy affects financial stability, we estimate the impact on the banking system (sum of all the banks' equities) caused by a full production disruption of a single firm, $j$, by computing the resulting supply chain  contagion and loan defaults. For the initial failure of firm $j$ we define single-firm shock vector 
\begin{equation}
    \psi^{\text{S},j}_i= 
    \begin{cases} 
  1 \qquad  \text{if} \quad i\neq j \\ 
  0 \qquad \text{if} \quad i=j \quad . \\ 
  \end{cases}
\end{equation}
The value $\psi^{S,j}_i\in \{0,1\}$ indicates the production level of firm $i$ under the initial shock scenario, $j$ (failure of firm $j$).\footnote{ The super script S indicates that the shock vector, $\psi^{\text{S},j}$, represents the failure of a \textit{single} firm $j$.}
The set $\Psi^\text{S} = \{\psi^{S,1}, \psi^{S,2}, \dots, \psi^{S,j}, \dots, \psi^{S,n} \}$, collects the $n$ possible shock scenarios,  corresponding to a  failure of a single firm.  
Note that we use the indices $i,j \in \{1,2,...,n\}$ to indicate firms in the SCN and $k\in \{1,2,...,m\}$ for banks. For every initial single firm failure, $\psi^{\text{S},j}$, we perform \textit{steps $1-3$} and receive the fraction of lost equity, $\mathcal{L}_k=\mathcal{L}_k (\psi^{\text{S},j})$, for every bank $k$. Based on the equity losses, $\mathcal{L}_k$, we compute the \textit{financial systemic risk index (FSRI)} of firm $j$,  as
\begin{equation}
    \rm{FSRI}_j \equiv \sum_{k=1}^m \frac{e_{k}}{\sum_{\ell=1}^m e_{\ell}} \min \left[1,\mathcal{L}_k \left(\psi^{\text{S},j} \right) \right] \quad .
    \label{FSRIdef}
\end{equation}
$\rm{FSRI}_j$ can be interpreted as the \textit{equity-weighted sum of losses suffered by individual banks in the network}. It is  the fraction of overall bank equity that is lost after the initial failure of a single firm propagates through the supply chain network, $W$,  causing firm insolvencies and loan write offs. It can be used to rank the companies operating  in the production network by their systemic relevance to financial stability. Note the difference to simply computing the DebtRank of companies in the credit network without their role in the supply chains, as was done in \cite{poledna2018identifying, Landaberry}.

Fig.\ref{twolayernetwork}, panel (c) shows for our example SCN how FSRI$_f$ of firm $f$ is computed as the weighted average of the direct and indirect losses shown in  panel (b). The weights are given by the relative equity size of every bank, i.e. $w_3 = w_4 = 0.25=1/m$, since we assume that all banks in our example have the same equity. We see that $\rm{FSRI}_f=0.15=\mathcal{L}(\psi)$, meaning that $15\%$ of the overall bank equity is at risk should firm $f$ fail. Similarly, if we simulate the initial failure of the other firms in the example SCN, we expect system-wide losses of $20\%,$ $15\%$, $5\%$, $5\%$, and $0\%$ for the shock vectors $\psi^{\text{S},c}$, $\psi^{\text{S},d}$, $\psi^{\text{S},b}$, $\psi^{\text{S},e}$, and $\psi^{\text{S},a}$, respectively. 

To highlight the effect from supply chain contagion, we distinguish between the \textit{direct} and \textit{indirect} components of FSRI
\begin{equation}
    \rm{FSRI}_j^{\rm{dir}} \equiv \sum_{k=1}^{m} \frac{e_{k}}{\sum_{l=1}^m e_{l}} \min\left[1,\mathcal{L}_k^{\rm{dir}} \left(\psi^{\text{S},j}\right)\right] \quad and \quad \rm{FSRI}_j^{\rm{indir}} = \rm{FSRI}_j-\rm{FSRI}_j^{\rm{dir}} \quad .
    \label{FSRIdir}
\end{equation}
For every firm the direct component  is proportional to the loans of that firm; the indirect FSRI is equal to the weighted sum of loans of firms that defaulted because of the operational failure of the firm $j$.

If  in Eqs. (\ref{lossesLambdak}) and (\ref{FSRIdef}) $\chi$ is  substituted by $\chi_{\textrm{eq}}$ or by $\chi_{\textrm{l}}$, we get the FSRI  triggered by equity or liquidity defaults,
\begin{equation}
    \textrm{FSRI}_{\textrm{eq,}j} \equiv \sum_{k=1}^m \frac{e_{k}}{\sum_{l=1}^m e_{l}} \textrm{min}\left[1,\mathcal{L}_k \left(\chi_{\textrm{eq}}(\psi^{\text{S},j})\right)\right] \quad {\rm and}
     \label{FSRIeq}  
\end{equation}
\begin{equation}
    \textrm{FSRI}_{\textrm{l,}j} \equiv \sum_{k=1}^m \frac{e_{k}}{\sum_{l=1}^m e_{l}} \textrm{min}\left[1,\mathcal{L}_k \left(\chi_{\textrm{l}}(\psi^{\text{S},j})\right)\right] \quad ,
     \label{FSRIl}  
\end{equation}
respectively.
\\

\textbf{ Estimating individual bank losses from supply chain contagion.}
To answer the second research question about how much the risk of individual banks is amplified by supply chain contagion, we need to investigate the exposure of  banks to losses transmitted by the supply chain network under more general initial shock scenarios. 
Typical measures used by banks to assess the riskiness of portfolios are  \textit{expected loss} (EL),  \textit{value at risk} (VaR) and  \textit{expected shortfall}  (ES) \citep{mcneil2015quantitative}. These measures are calculated from a \textit{loss distribution} \citep{mcneil2015quantitative}. Here we estimate the loss distribution for each bank with a Monte Carlos simulation that generates initial shock scenarios --- representing the simultaneous failure of multiple firms --- to the supply chain network. In this way we generate different stress scenarios for the economy and simulate the corresponding losses for each bank.

Specifically, we simulate $10,000$ initial shocks, $\psi^{\text{M},\ell}$,  where multiple firms suffer a full production disruption. 
The value $\psi^{M,\ell}_i\in \{0,1\}$ indicates the production level of firm $i$ under the initial shock scenario, $\ell$.  
We use the index $\ell \in \{1,2,...,10000\}$ to distinguish the 10,000 scenarios. 
Every shock scenario vector, $\psi^{\text{M},\ell}$, is created by drawing from a $n$-dimensional multivariate Bernoulli random variable, where the success probabilities correspond to the $n$-dimensional vector of default probabilities (PDs) of firms, i.e. 
%
$$\psi_i^{\text{M},\ell} \sim \rm{Bernoulli}(\text{pd}_i) \quad ,$$ 
where $\text{pd}_i$ is the default probability of firm $i$. We set the initial production level of firm $i$ to $\psi^{\text{M},\ell}_i=0$ in case of a success and to $1$, otherwise.
The PDs are estimated by the Central Bank of Hungary, for details see \ref{appendixB}.
Note that due to data limitations we draw the defaults of firms independently, i.e. we neglect systematic or correlated events like a large crises affecting specific industry sectors.  
We denote the set of initial shock vectors by $\Psi^\text{M} = \{\psi^{\text{M},1}, \psi^{\text{M},2}, \dots , \psi^{\text{M},\ell}, \dots \psi^{\text{M},10000} \} $ \footnote{The super script M indicates that the shock vector, $\psi^{\text{M},\ell}$, represents the simultaneous failure of \textit{multiple} firms.}.

As before, contagion leads to direct and indirect losses for banks. 
For each bank $k$, we compute the two loss distributions consisting of its \textit{contagion-adjusted} equity losses, $\big( \mathcal{L}_k(\psi^{\text{M},1}), \mathcal{L}_k(\psi^{\text{M},2}), \dots, \mathcal{L}_k(\psi^{\text{M},10000})$ and its \textit{direct} equity losses  $\big(\mathcal{L}_k^{\textrm{dir}}(\psi^{\text{M},1}),  \mathcal{L}_k^{\textrm{dir}}(\psi^{\text{M},2}),  \dots,   \mathcal{L}_k^{\textrm{dir}}(\psi^{\text{M},10000})  \big)$,  corresponding to the 10,000 shock scenarios, $\Psi^\text{M}$.
%
%
Similarly, $\mathcal{L}(\psi^{\text{M},\ell} )$ is used to obtain the loss distribution for the entire banking system. 
For a given shock vector $\psi^{\text{M},\ell}$, the losses $\mathcal{L}_k(\psi^{\text{M},\ell})$ and $\mathcal{L}_k^{\rm{dir}}(\psi^{\text{M},\ell})$ are obtained by performing 
\begin{itemize}
    \item Step \hyperref[step1]{1}  for $\psi= \psi^{\text{M},\ell}$ and yielding the production losses of firms, $h(T)$;
    \item Step \hyperref[step2]{2} yielding the vector of defaulted firms, $\chi(\psi^{\text{M},\ell})$;
    \item Step \hyperref[step3]{3} yielding the equity losses of banks, $\mathcal{L}_k(\psi^{\text{M},\ell})$, for $\chi(\psi^{\text{M},\ell})$;
    and $\mathcal{L}_k^{\rm{dir}}(\psi^{\text{M},\ell})$ for $\chi^{\textrm{dir}}(\psi^{\text{M},\ell})$ \footnote{Finding $\mathcal{L}_k^{\textrm{dir}}(\psi^{\text{M},\ell})$ is equivalent to a procedure in which step 1, i.e. contagion, is skipped and $h(T)$ is set to be equal to the initial shock, $\psi^{\text{M},\ell}$. Then, it is used in step 2 and $\chi^{\textrm{dir}}$ is obtained from Eq. (\ref{chi}).}.
\end{itemize}
The procedure is repeated for every $\ell \in \{1,2,...,10000\}$.

For the VaR and ES we choose the $95\%$ threshold. To compare these risk measures for the direct and the contagion-adjusted loss distributions, we use the \textit{risk amplification factor}, $\rho_k^X$ for bank $k$, where $X$ stands for EL, VaR, or ES. We define it as
\begin{equation}
    \rho_k^\text{EL}  \equiv \frac{\Tilde{\mu}^\text{EL}_k}{\mu_k^\text{EL}} \qquad , \qquad 
    \rho_k^\text{VaR} \equiv \frac{\Tilde{\mu}^\text{VaR}_k}{\mu_k^\text{VaR}} \qquad , \qquad  
    \rho_k^\text{ES} \equiv \frac{\Tilde{\mu}^\text{ES}_k}{\mu_k^\text{ES}} \qquad ,
    \label{riskampl}
\end{equation}
where $\mu^X_k$ is the value calculated from the direct- and $\Tilde{\mu}^X_k$ from the contagion-adjusted loss distribution of bank $k$. We denote the averages over all banks by $\rho^{\textrm{EL}}$, $\rho^{\textrm{VaR}}$, and $\rho^{\textrm{ES}}$.

\section{Data} \label{data}

We calibrate the model to a unique real-world firm-level data set composed of 4 distinct micro-data sets that are available within the \textit{Central Bank of Hungary}. These are the
\begin{enumerate}
    \item VAT based supplier-buyer relationships between practically all firms in Hungary;
    \item balance sheets and income statements of firms;
    \item loans of banks to firms;
    \item CET1 equity of banks.
\end{enumerate}

The first dataset is used to reconstruct the supply-chain network, $W$. It is based on the 2019 value added tax (VAT) reports, reflecting any purchase between two firms exceeding 100,000 HUF tax content (approximately $250$ Euros) made in Hungary. The resulting network consists of $243,339$ anonymized Hungarian companies. The link weights correspond to the monetary value of transactions (price times quantity). We filter the links such that they contain only relationships that are stable. We include links that did occur in at least 2 transactions in two different quarters of the year 2019. $52\%$ of links are stable in this sense and they cover $93\%$ of the traded volume in the network. For a more detailed description of the data set based of the year 2017, see \citep{borsos2020unfolding, diem2022quantifying}. 

The second dataset consists of the financial statements of every firm in the supply chain network. The variables obtained from the balance sheets and income statements, see Table \ref{table:IncStBalSh}, are used to calculate the equity and liquidity buffers of firms, as well as their profits. For every firm, we have an estimate of its default probability obtained from a model developed at the Central Bank of Hungary, see \cite{burger2022defaulting} and \ref{appendixB} for more details.

The third dataset, the credit registry, is needed to determine the firm-bank loans, $B$. It consists of the overall exposures of $40,043$ firms in the supply chain network, $W$, to $269$ banks in Hungary. The remaining $203,296$ companies do not have reported bank loans in Hungary. We remove $247$ small-scale banks for which we do not have \textit{Common Equity Tier 1 (CET1)} values (contained in the fourth data set) available, arriving at $27$ banks that cover $84\%$ of the loan volume of the original dataset with $269$ banks. Hence, $B$ is a matrix of size $n\times m$ with $35,609$ non-zero inputs (multi-layer links), with $n=243,339$ and $m=27$. $32,256$ firms have loans from one of the $27$ banks.

Based on this information the model is fully data-driven, meaning that there are no free parameters. The only modelling choices concern the, set-up of the SCN shock propagation (taken from \cite{diem2022quantifying, diem2023estimating}),  assumptions of how production shocks affect the financial health of a firm, i.e. that revenues and material costs can be reduced proportionally to production losses, the choice that a decrease of $100\%$ in equity or liquidity cause bankruptcy, and the assumption that LGD is equal to $100\%$, see more in Discussion \ref{discussion}.

\section{Results}  \label{results}
 
\begin{figure}[tbh]
 \centering
    \includegraphics[scale=.06, keepaspectratio]{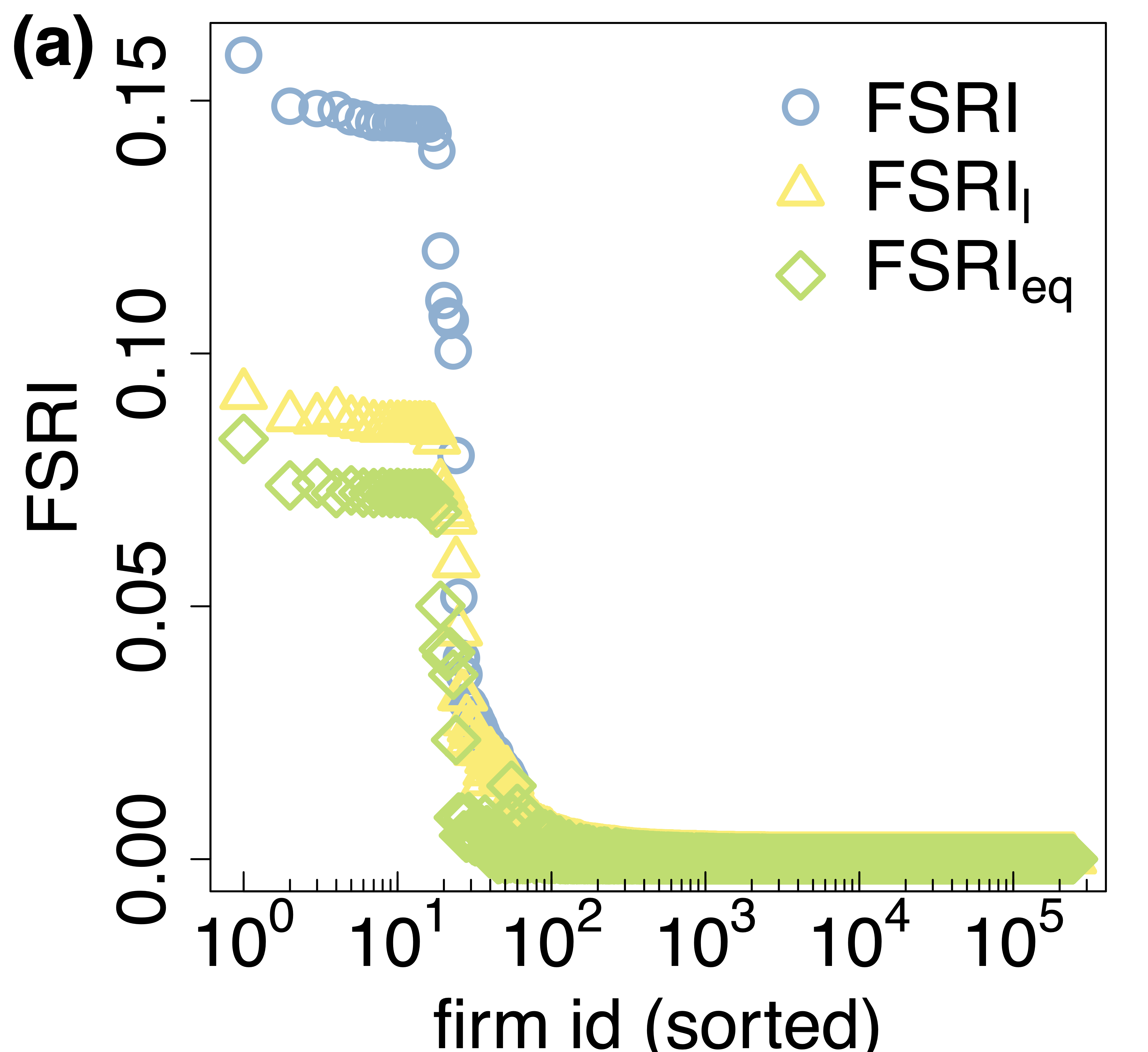}
    \includegraphics[scale=.06, keepaspectratio]{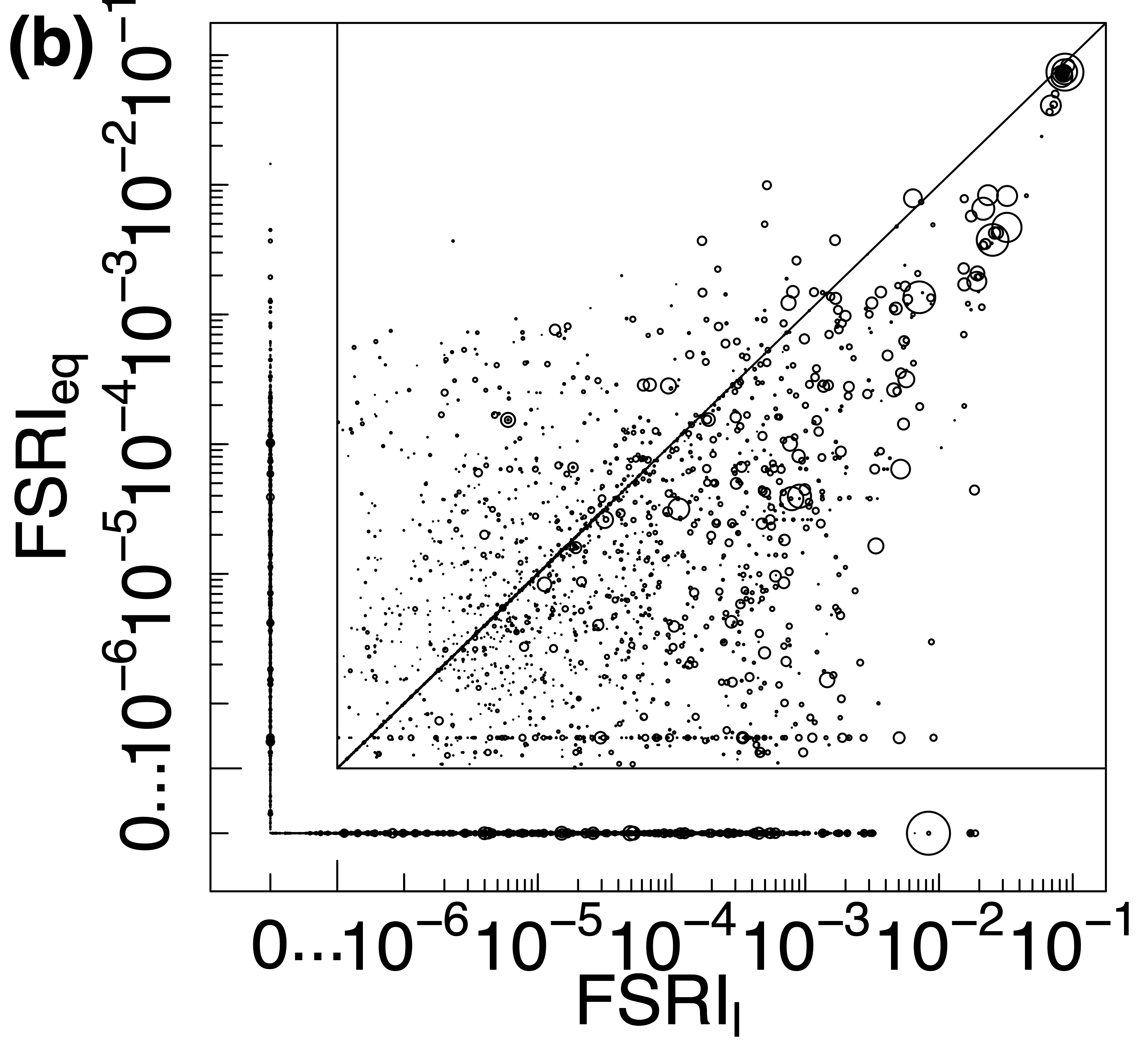}
 \caption{\textbf{Financial systemic risk index (FSRI) of firms in the Hungarian SCN}.  \textbf{(a)} Log-linear scatter plot of FSRI of firms sorted in decreasing order of $\textrm{FSRI}$ (blue circles). The $x$-axis refers to firms' FSRI-rank from $1$ to $243,339$, the $y$-axis represents the FSRI values of those firms. FSRI is shown for two causes of insolvency independently: $\textrm{FSRI}_{\textrm{l}}$ (yellow triangles) - due to negative short-term liquidity and $\textrm{FSRI}_{\textrm{eq}}$ (green diamonds) due to negative equity. $\textrm{FSRI}$ (blue circles) is due to either one of them. 
 All three distributions are sorted in the same way with respect to FSRI.
 Out of $243,339$ firms in the SCN about $100$ are systemically risky with $23$ (first 23 from left) having very large systemic risk, $10-16\%$ of the overall banks' equity is lost if any of them fails. These 23 companies are the same for all three  distributions. \textbf{(b)}    Log-log (plus a zero-section) scatter plot  of the $\textrm{FSRI}_{\textrm{l}}$ against the $\textrm{FSRI}_{\textrm{eq}}$, where firms with high values of $\textrm{FSRI}_{\textrm{l}}$  also have the highest values of $\textrm{FSRI}_{\textrm{eq}}$; see the upper right corner. The size of a bubble is proportional to the out-strength (sales) of a firm. The plot reveals that large systemic risk is not necessarily proportional to the sales of a firm.} 
 \label{FSRI}
\end{figure}

\begin{figure}[tbh]
	\centering
	\includegraphics[scale=.06, keepaspectratio]{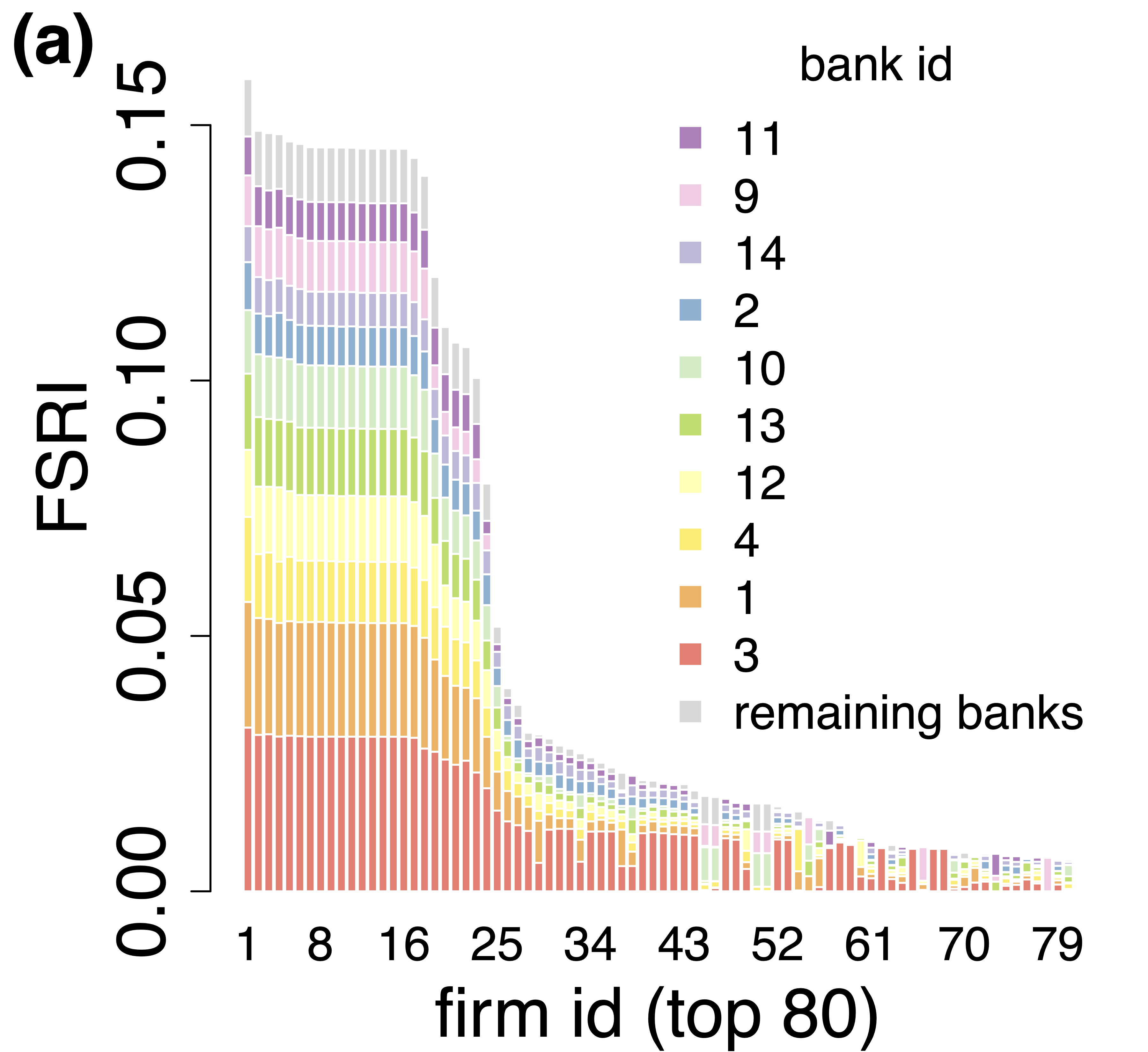}
	\includegraphics[scale=.06, keepaspectratio]{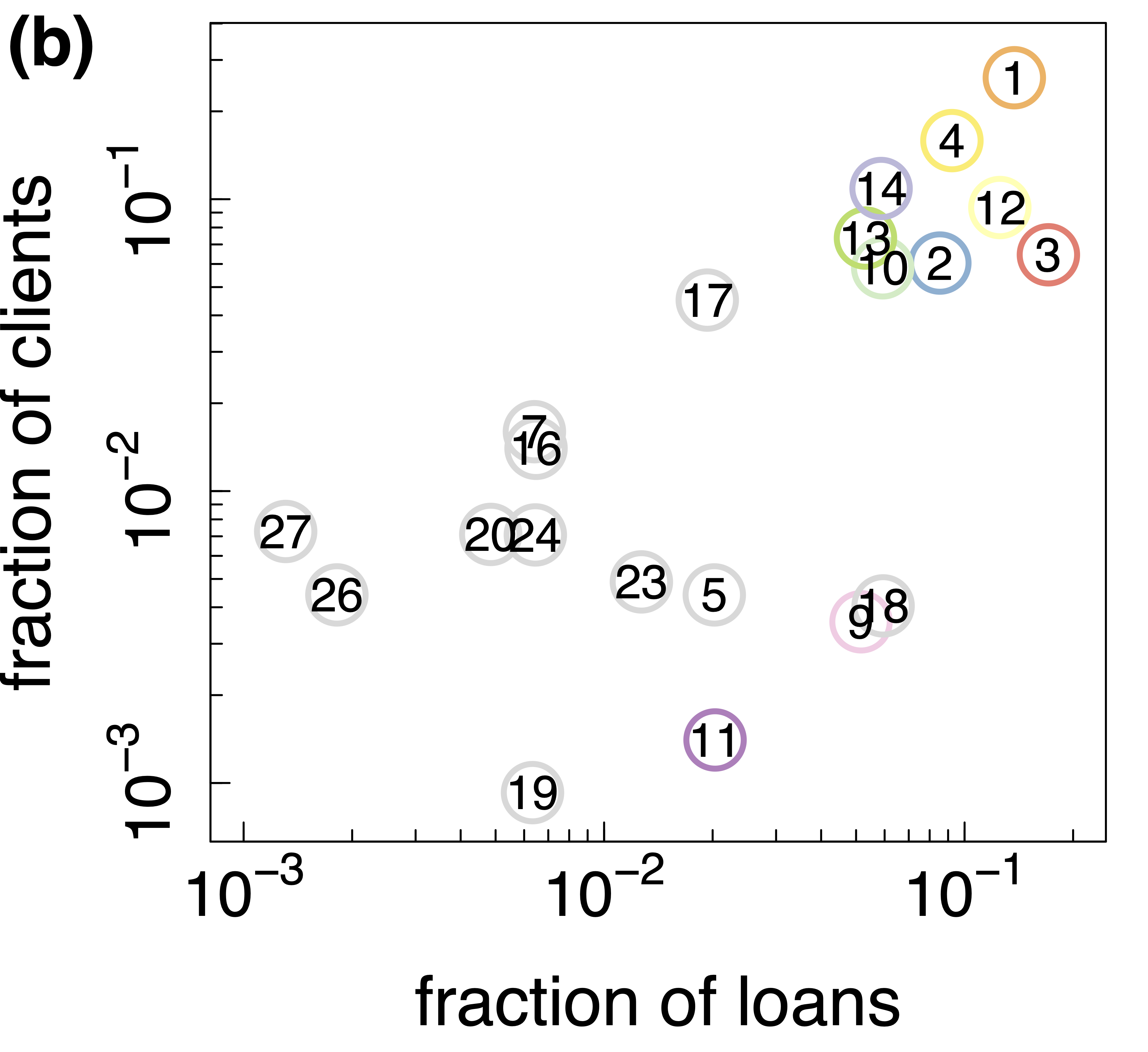}
 	\caption{\textbf{FSRI by losses for each bank.}  \textbf{(a)} FSRI of the first $80$ firms disaggregated to banks. 
	Losses are suffered mainly by $10$ banks, namely 1, 2, 3, 4, 9, 10, 11, 12, 13, 14. \textbf{(b)}  log-log scatter plot shows the fractions 	of loans versus bank creditors (we omit 6 banks with fractions below $0.1\%$). It is visible that bank 3 has the largest amount of loans, 	$17\%$, and bank 1 has the biggest proportion of clients, $26\%$. Banks in the upper right corner (1, 2, 3, 4, 10, 12, 13, 14) suffer 		substantial  losses from the riskiest firms in the bar-plot.
  }
\label{FSRIbanks}       
\end{figure}

We compute the financial systemic risk index (FSRI) (\ref{FSRIdef}) for each of the $n=243,339$ companies contained in the Hungarian supply chain network. The rank-sorted distribution of the FSRI values (blue circles) is shown in Fig.\ref{FSRI}(a) in log-linear scale,  where the $x$-axis denotes firm ranks and the $y$-axis their respective FSRI values. The firm with the highest financial systemic risk is to the very left and the lowest systemic risk firms to the very right. The riskiest firm has an FSRI value of slightly above $15\%$, meaning that its failure and the ensuing supply chain disruption cascade would lead to loan write offs equivalent to approximately $16\%$ of the overall CET1 capital of the $27$ Hungarian banks in the sample (see Fig.\ref{FSRIbanks}(a) for how banks are affected individually). A group of $17$ firms form a plateau with very similar FSRI values slightly below $15\%$. This high-risk plateau is followed by a sharp drop in FSRI values of $7$ more firms with values above $5\%$ and $35$ firms with values between $5\%$ and $1\%$. In total, $308$ firms have FSRI values higher than $0.1\%$. To a large extent the relatively high risk of these firms is driven by the substantial supply chain cascades they trigger (indirect losses). The firms in the high FSRI plateau cause similar supply chain cascades, i.e., they affect mostly the same firms in response to their failure and hence also their financial impacts on the firms in the network is similar, finally causing similar losses to banks' equity levels. This pattern of similar supply chain cascades can be explained by the observed  plateau firms, forming a tightly knit network (systemic core of the economy) of highly risky supply relations \citep{diem2022quantifying}. 
These companies predominantly belong to the energy, communication, and transportation sectors, a few belong to waste collection and manufacture of basic chemicals. These sectors have been identified as essential to many other industries in the survey conducted by \citep{pichler2022forecasting}, which is used as input to the supply chain shock propagation algorithm, see \ref{appendixA}.

Next we distinguish between the bank capital losses that originate from insolvencies due to a lack of liquidity or a lack of equity. The financial systemic risk index, $\rm{FSRI}_{\rm{eq}}$ (\ref{FSRIeq}), solely based on firm insolvencies caused by their equity turning negative after supply-chain shock propagation is shown by green diamonds. Similarly, the financial systemic risk index, $\rm{FSRI}_{\rm{l}}$ (\ref{FSRIl}), solely based on firm insolvencies caused by their liquidity turning negative after supply-chain shock propagation is shown by yellow triangles. As expected, $\textrm{FSRI}_{\textrm{eq}}$ and $\textrm{FSRI}_{\textrm{l}}$ are consistently smaller than $\textrm{FSRI}$, since the insolvency criterion for FSRI encompasses both, equity and liquidity based insolvencies. Remarkably, the high-risk plateaus of $\textrm{FSRI}_{\textrm{eq}}$ and $\textrm{FSRI}_{\textrm{l}}$ show similar heights, of $7\%$ and $8\%$, respectively. This indicates that as a result of the largest supply-chain disruption cascades most firms only suffer from either their equity \textit{or} their liquidity turning negative; only a few firms suffer from both at the same time. In the $\textrm{FSRI}_{\textrm{eq}}$ profile distribution there are $18$ firms with risk over $6\%$, additional $6$ firms in the interval $3-6\%$, and in total $107$ firms with risk higher than $0.1\%$. Similarly, in the $\textrm{FSRI}_{\textrm{l}}$ profile distribution there are $18$ firms with risk over $7\%$, additional $6$ firms in the interval $5-7\%$, and a total of $265$ firms exceeding the risk of $0.1\%$. 

To investigate the relation between liquidity- and equity based insolvencies more closely, we provide a scatter plot in log-log scale in Fig.\ref{FSRI}(b). Since, many firms only cause either liquidity or equity based insolvencies, we plot two additional segments that would not be visible on the log-log scale. The first one shows the $\textrm{FSRI}_{\textrm{eq}}$ values if  $\textrm{FSRI}_{\textrm{l}} = 0$  and the second segment shows the $\textrm{FSRI}_{\textrm{l}}$ values if $\textrm{FSRI}_{\textrm{eq}} = 0$. Bubble size is proportional to the out-strength (sales to other firms) of the respective firm (see \ref{appendixA}). Firms in the FSRI plateau cause both, large amounts of equity- and liquidity based losses (top right, close to the diagonal). One observes that these are caused by large (large bubbles) and small firms (small bubbles). For small values $\textrm{FSRI}_{\textrm{l}}$ and $\textrm{FSRI}_{\textrm{eq}}$ seem uncorrelated. There are $22,600$ firms with non-zero $\textrm{FSRI}$ values, $16,208$ for $\textrm{FSRI}_{\textrm{l}}$ and $12,301$ for $\textrm{FSRI}_{\textrm{eq}}$. The low number of positive FSRI values can be explained by two factors. First, even though economic systemic risk  (ESRI) --- measuring the fraction of the production that is lost in the supply chain network after the failure of a firm --- values decay as power law (see \cite{diemsupplementary} Fig. S4), only a few firms cause large supply-chain cascades that can affect the financial conditions of other firms severely enough to cause a default. Second, in the data set $13\%$ of firms have loans, i.e., only $32\,256$ firms out of $243,339$ have loans in one of the $27$ banks. Hence, indirect losses to banks are only caused when supply chain contagion leads to default of at least of these firms.

\begin{figure}[htb]
 \centering
    \includegraphics[scale=.06, keepaspectratio]{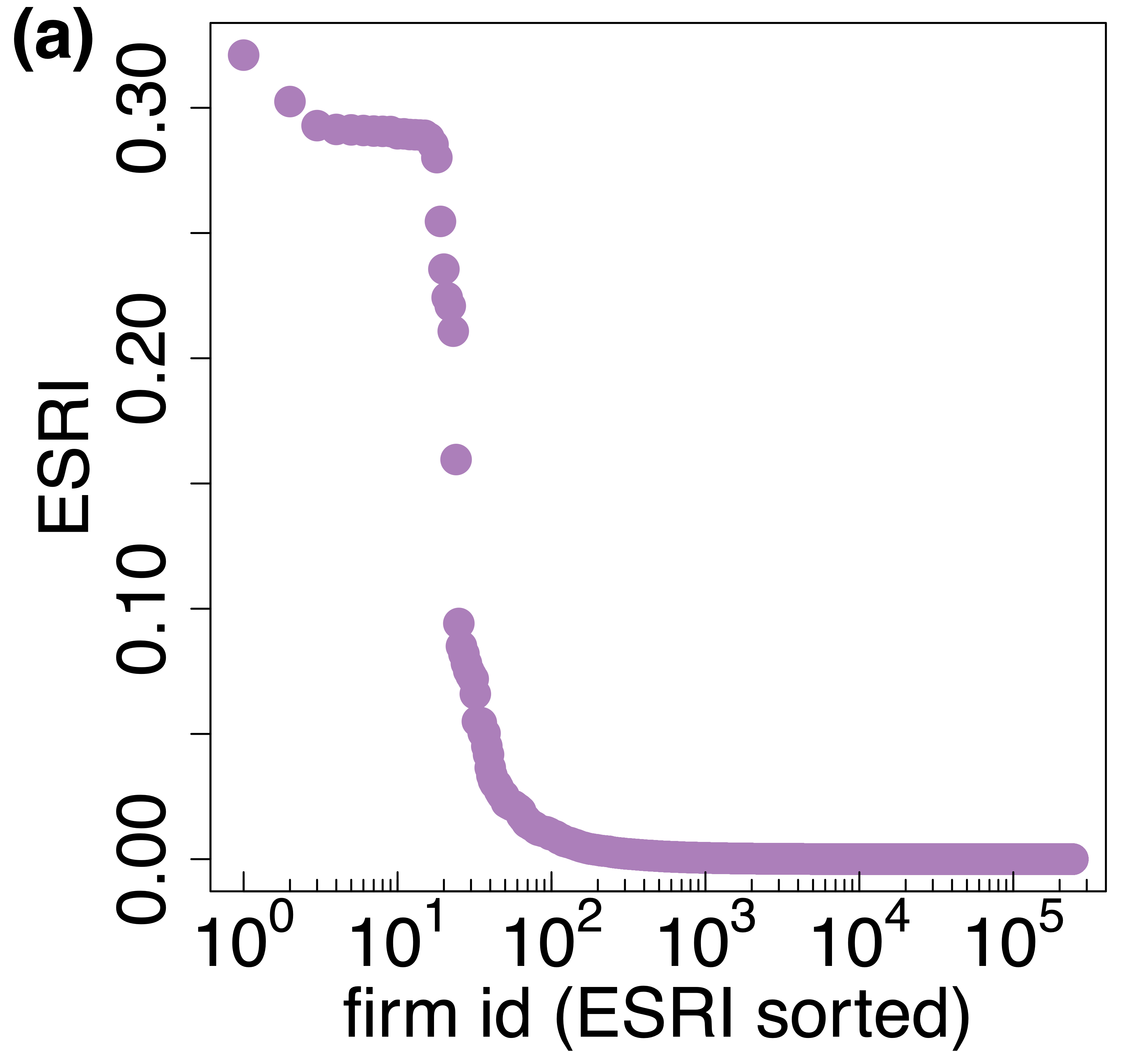}
    \includegraphics[scale=.06, keepaspectratio]{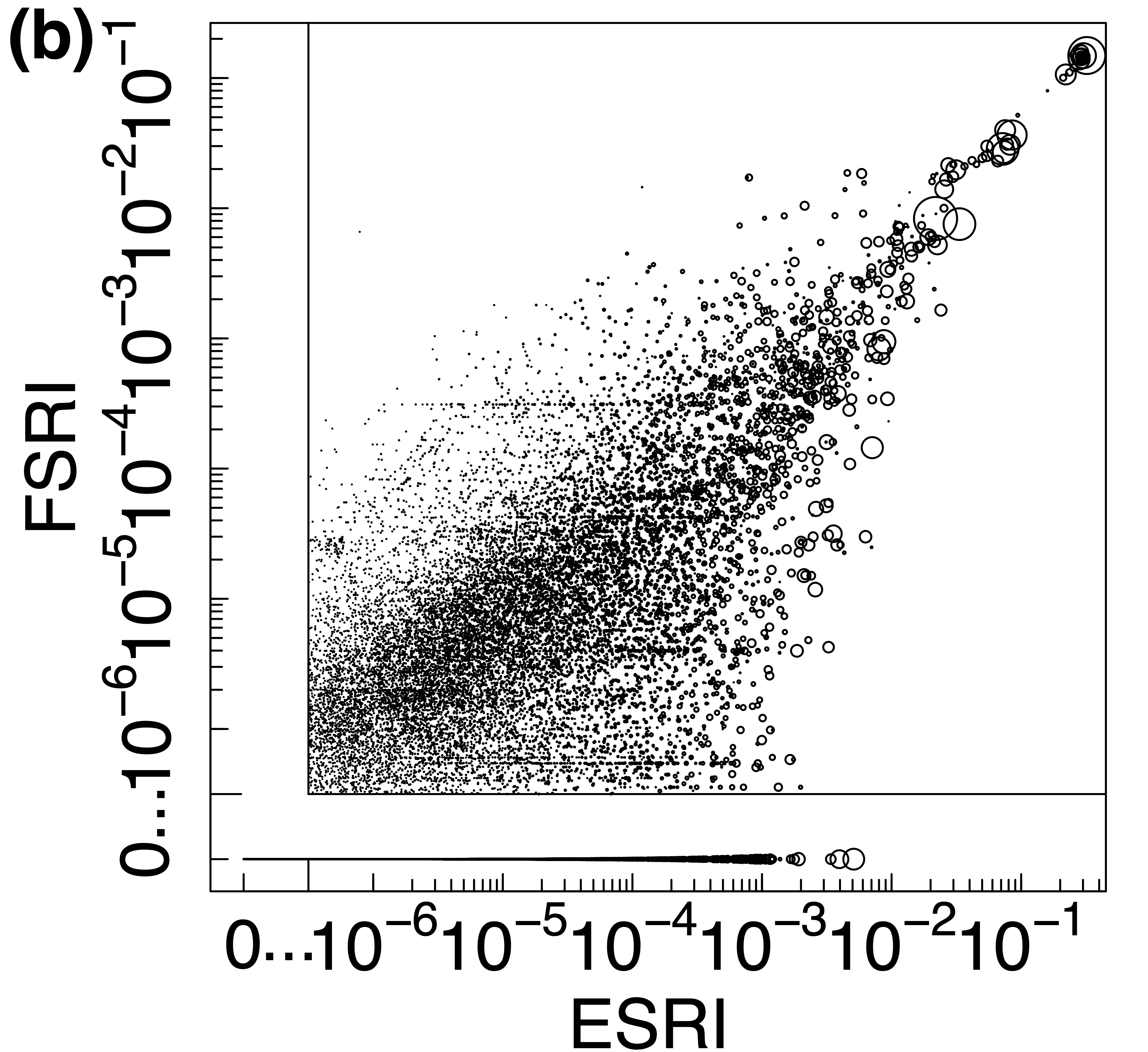}
 \caption{\textbf{Comparison of the economic systemic risk index (ESRI) with FSRI.} \textbf{(a)}  log-liner plot of firms' ESRI (purple dots) rank-sorted in decreasing order from $1$ to $243,339$. The ESRI of a firm is the fraction of lost production in the supply chain network after an initial disruption of that particular firm and the ensuing shock propagation. The distribution of ESRI is qualitatively very similar to FSRI. The $24$  most systemically risky companies lead to production losses between 10\% and 32\%. These are mostly the same companies appearing in the FSRI profile in Fig.\ref{FSRI}. This is evident also in the scatter plot in panel \textbf{(b)}. Bubble size is the out strength (sales). Most risky firms have high out strength, but there are firms with small out strength and high systemic financial and/or economic risks. 
 } 
 \label{ESRIplot}
\end{figure}

In Fig.\ref{ESRIplot} we study the relation between FSRI and ESRI. The later quantifies the size of the supply-chain disruption cascade (in the real economy) in terms of the fraction of the production network's total output that is affected in case of a firm's failure; for details on ESRI and its calculation see \ref{appendixA}, Eq. (\ref{esri}) and \cite{diem2022quantifying}.
Figure \ref{ESRIplot}(a) shows the characteristic shape (high-risk plateau, steep (power-law) drop, and many small-risk firms) of the ESRI distribution (purple dots) as observed in \citep{diem2022quantifying}. These features obviously carry over to the FSRI values (blue circles)  as shown in Fig.\ref{FSRI}(a). The $24$  most systemically risky companies lead to production losses between 10\% and 32\%  in case of their failure. These firms coincide with the high-risk firms in the FSRI plateau. This is visible in the log-log (plus zero section) scatter plot in Fig.\ref{ESRIplot}(b) with ESRI on the x-axis and FSRI on the y-axis. Bubble size is again proportional to out strength. We see that for the highest values of ESRI and FSRI are strongly correlated. The overall correlation coefficient of non-zero values is about $0.5$ on the log-log scale. 
There are many firms with an FSRI equal to $0$, but relatively large ESRI values up to around $0.5\%$. The likely reason is that firms'  supply-chain cascades, even though large, do not affect the financial conditions of other firms severely enough to cause their default or cause defaults to firms without loans.  Again, many firms do not have bank loans; see Section \ref{data}).

\begin{figure}[tbh]
	\centering
	\includegraphics[scale=.06, keepaspectratio]{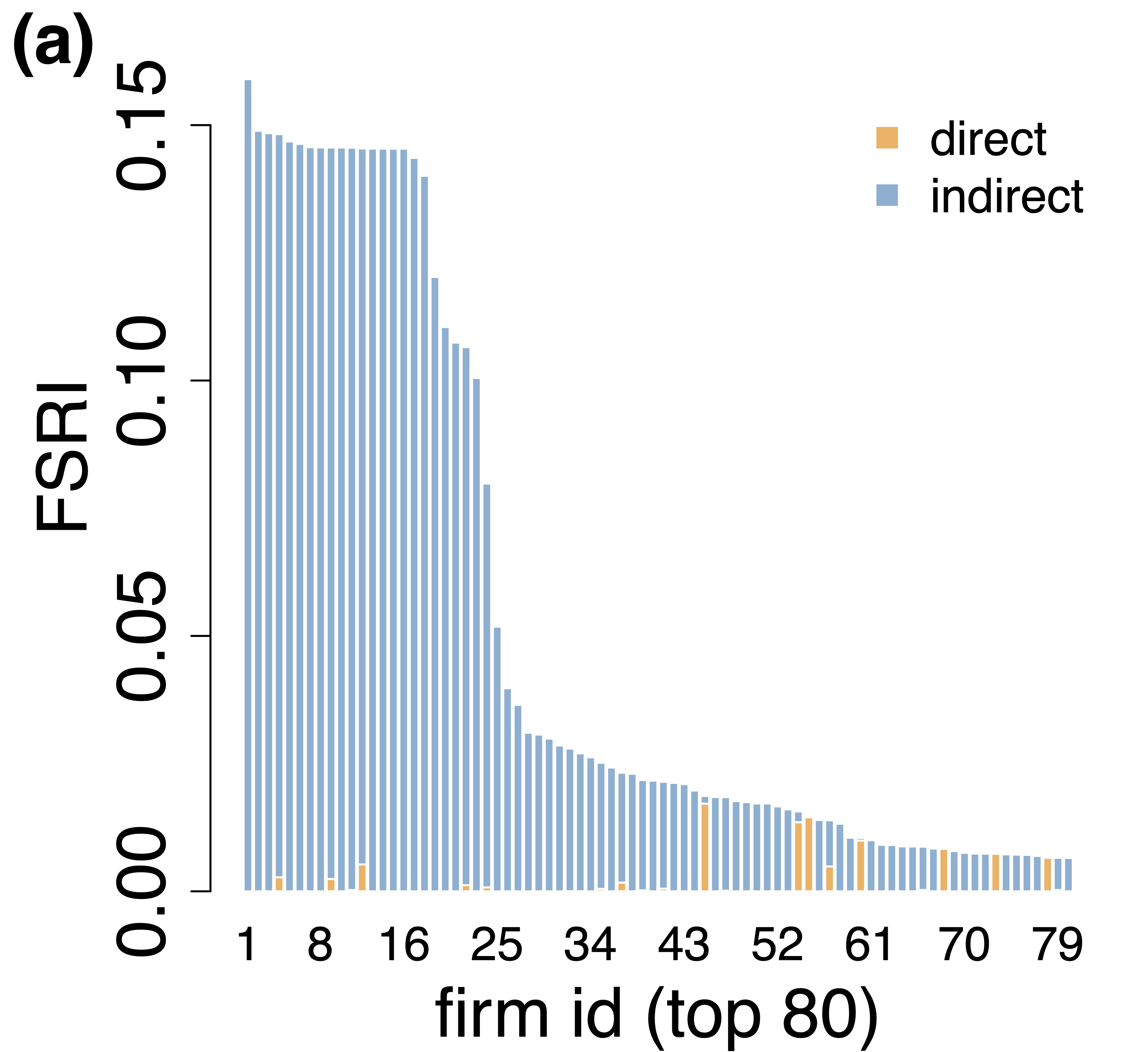}
	\includegraphics[scale=.06, keepaspectratio]{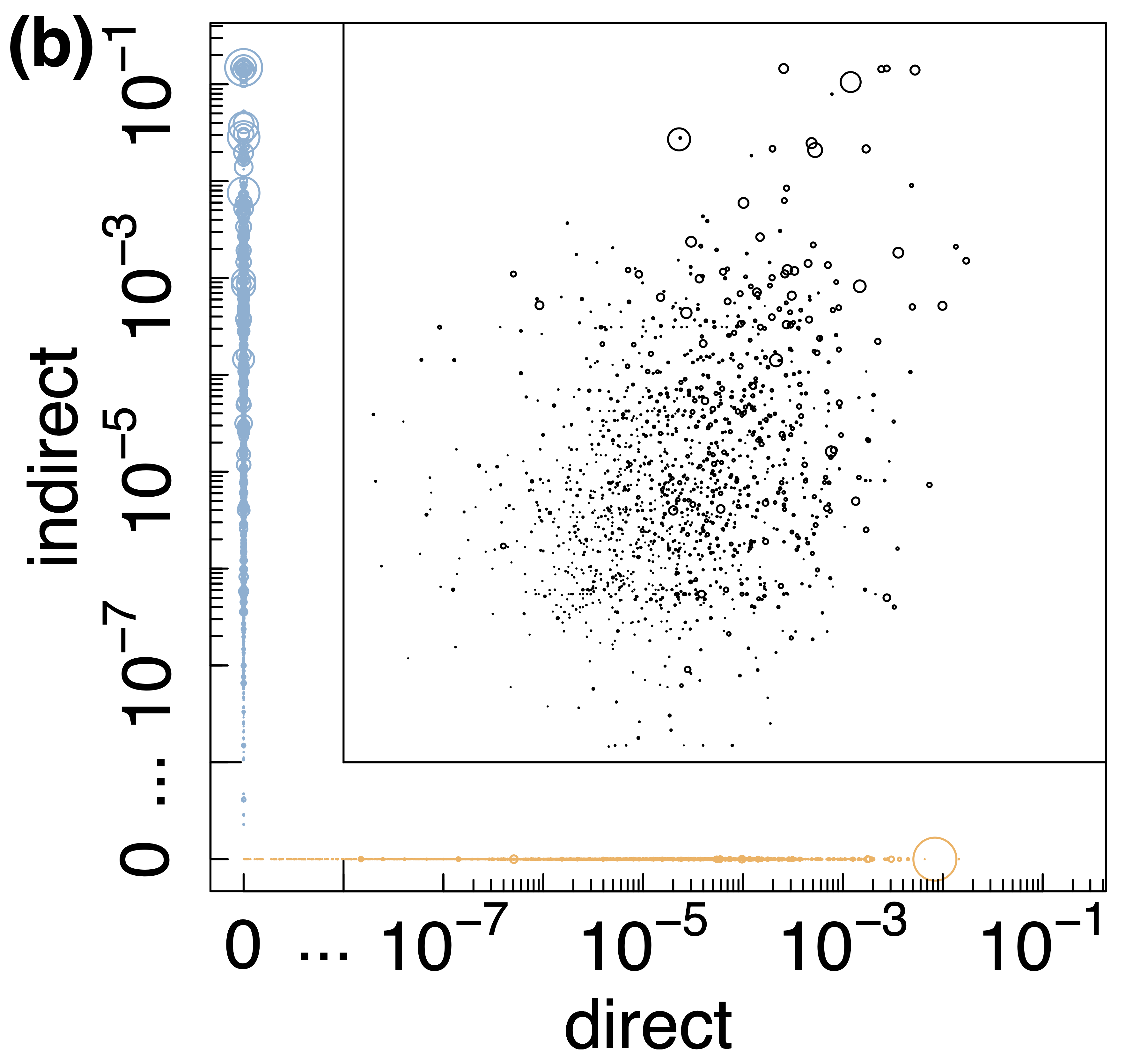}
 	\caption{\textbf{Direct and indirect components of FSRI.} \textbf{(a)} shows the FSRI profile for the first $80$ companies, decomposed 	into the direct and indirect contributions. 
	The direct share $\textrm{FSRI}_j^{\textrm{dir}}$ (orange bar) in the FSRI is proportional to company's loans. The indirect share $\textrm{FSRI}_j^{\textrm{indir}}$ (blue bar) is proportional to loans of firms that default after $j$ triggers a cascade. It is obvious, that the riskiest firms cause substantial financial losses indirectly through spreading shocks along the SCN. \textbf{(b)}  log-log (plus a zero section) scatter plot of direct and indirect FSRI shares of all firms.  Circle size is the out-strength. Most large firms typically cause indirect effects.}
	\label{FSRIdirectindirect}       
\end{figure}

We next investigate the relation between losses caused directly by firms defaulting on their own loans and indirect loan defaults caused by secondary supply chain disruptions. Figure \ref{FSRIdirectindirect}(a), disaggregates the FSRI values into direct and indirect losses for those 80 firms that cause the largest bank equity losses. The direct share $\rm{FSRI}_j^{\rm{dir}}$ (orange bar) of firm $j$ (see Eq. (\ref{FSRIdir})) is simply the size of its defaulted loans divided by overall banks' equities. Naturally most loans are very small with respect to  total bank equity. The indirect share $\rm{FSRI}_j^{\rm{indir}}$ (blue bar) defined in Eq. (\ref{FSRIdir}) is proportional to loans that default after $j$ triggered a supply-chain cascade. The overwhelmingly blue color in the bar-plot indicates that the highest losses to overall bank equity are caused by supply chain shock propagation and not by the size of the firms' own loans, see also Fig.\ref{FSRIoutstrloans}. A direct loss can be caused only by firms that have loans. If a firm without a loan initially fails and defaults it won't cause a direct loss. For instance, out of the $80$ firms in Fig.\ref{FSRIdirectindirect}(a) $39$ have loans and $45$ default. The two sets intersect for $23$ firms, whereas the remaining $16$ firms with loans do not default because of their strong equity and liquidity buffers, i.e., they cause only indirect losses. 

In Fig.\ref{FSRIdirectindirect}(b) we show the relation of direct versus indirect losses with a log-log (plus a zero section) scatter plot. In more detail,
$16,549$ companies have only positive direct effects, i.e $\rm{FSRI}_j=\rm{FSRI}_j^{\rm{dir}}$. $7,483$ firms (blue circles)  have only positive indirect effects, i.e. $\rm{FSRI}_j=\rm{FSRI}_j^{\rm{indir}}$. The remaining $1,445$ firms (black circles) have non-zero direct as well as indirect effects, i.e. $\rm{FSRI}_j=\rm{FSRI}_j^{\rm{dir}}+\rm{FSRI}_j^{\rm{indir}}$. Most of the biggest companies, with respect to out strength (bubble size) cause zero direct losses (blue circles). Nevertheless, in Fig.\ref{FSRIdirectindirect}(a) we see some firms with big loans causing small indirect effect. In particular, firms $45$, $54$, or $55$ (showing the highest orange bars) have loans size around $1.5\%$ of total banks' equities each, and belong to the top $1\%$ based on their revenue and sale sizes.

Having discussed the impact on a system-wide bank capital level, we now turn to how banks are affected individually after initial failures of the $80$ riskiest firms. We disaggregate the FSRI profile to the bank-level in Fig.\ref{FSRIbanks}(a). The bar-plot shows the contributions  from Eq. (\ref{FSRIdef}), the losses of $10$ banks are shown with different colors, the relative compound losses of the remaining $17$ banks  in grey. Again, since the failure of the top FSRI firms trigger highly similar supply chain cascades, they trigger highly similar financial losses. This is reflected by the fact that banks are affected in a very similar ways by these 17 initial firm-failures. However, in general not always the same banks are affected. Different banks have different business models and different client structures. Some banks have large corporate loan portfolios relative to their equity, others have a stronger focus on retail- and mortgage lending. Further, banks can have a strong focus on certain industry sectors such as agriculture. Therefore, some firms cause supply chain cascades that affect specific banks disproportionately. The log-log scatter plot in Fig.\ref{FSRIbanks}(b) shows the fractions of loans ($x$-axis) against the fractions of clients ($y$-axis) of banks in the financial layer (we skip 6 banks with fractions below $0.1\%$). Bank 3 has the biggest amount of loans; bank 1 has the biggest proportion of clients. The bar-plot reveals that losses are suffered  mainly by the $10$ banks 1, 2, 3, 4, 9, 10, 11, 12, 13, 14, covering $74\%$ of the entire equity in the financial layer. Those banks hold $93\%$ of the clients that have $85\%$ of loans in those banks. Banks 3, 1, 4, 12, 13 cover $53\%$ of total equity,  hold $65\%$ of clients having $58\%$ of loaned money in the system.

\begin{figure}[tbh]
	\centering
	\includegraphics[scale=.06, keepaspectratio]{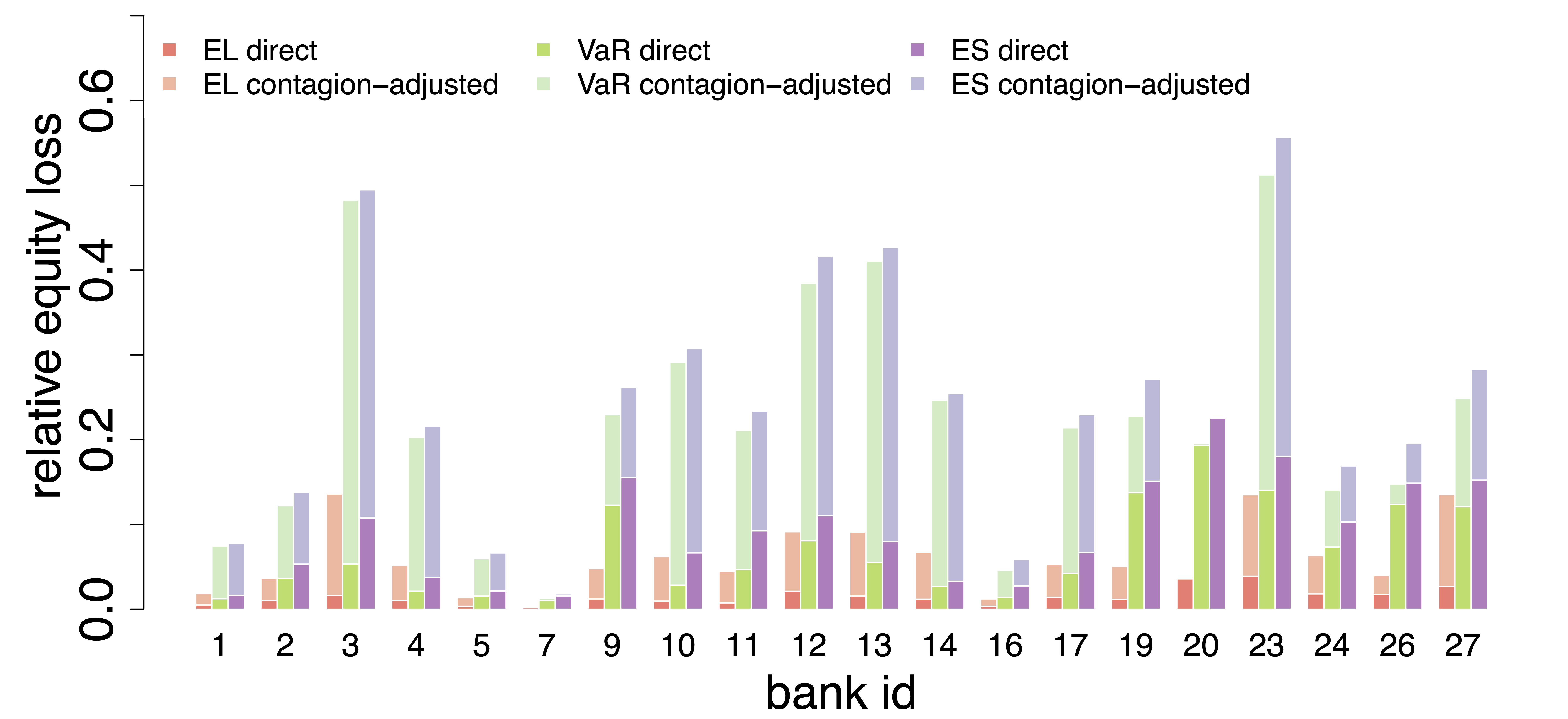}
 	\caption{\textbf{Expected losses, value at risk, and expected shortfall of banks including only direct and contagion-adjusted equity losses of banks}. Indices of banks (x-axis) are the same as in Fig.\ref{FSRIbanks}. We omit banks 6, 8, 15, 18, 21, 22, and 25 that have less than $30$ clients. The loss distributions are generated from relative equity losses, $\mathcal{L}_k(\Psi^M)$ from Eq. (\ref{lossesLambdak}), for 10,000 shock scenarios. The risk measures EL, VaR, and ES are shown in red, green and purple, respectively; direct losses in dark shades, contagion-adjusted losses in light shades. The bar-plot is not stacked, dark bars are plotted in front of light bars. Taking into account the exposure to the SCN massively amplifies risks for all banks. The average risk amplification factors are $\rho^{\textrm{EL}}=4.3$, $\rho^{\textrm{VaR}}=4.5$, and $\rho^{\textrm{ES}}=3.2$.
 	}
	\label{barplot_MC}       
\end{figure}

We next focus on how the credit risk exposure of individual banks is amplified in the presence of supply chain contagion. We measure the amplification by calculating the EL, VaR, and ES for the simulated losses of each banks' commercial loan portfolio, once without supply chain contagion (direct losses only) and with supply chain contagion. We calculate the direct losses, $\mathcal{L}_k^{\textrm{dir}}(\psi^{\text{M},\ell})$, and the supply chain contagion adjusted ones, $\mathcal{L}_k(\psi^{\text{M},\ell})$ with a Monte Carlo simulation for 10,000 initial shock scenarios, $\ell \in\{1,...,10\textrm{ }000\}$, as described in Section \ref{method}, step \hyperref[step3]{3}. EL is the mean of the losses,  VaR$_{0.95}$ is the 95\% quantile of losses, and the ES$_{0.95}$ is the average over the 500 largest losses.
Figure \ref{barplot_MC} shows the EL (red), VaR (green), and the ES (purple) for every bank from the direct (dark shade) and contagion-adjusted (light shade) loss distributions. Indices of banks are the same as in Fig.\ref{FSRIbanks}. We skip banks 6, 8, 15, 18, 21, 22, and 25 that have less than $30$ clients. The y-axis shows the value of the respective risk measures.
Note that the bars are not stacked (dark bars are always plotted in front of light bars) meaning that the height of the light shaded bars is the overall value of the risk measure with supply chain contagion. The height of the light bars minus the height of the dark bars gives the additional risk from contagion (marginal systemic risk). The plot shows that EL, VaR, and ES are substantially lower when the supply chain is not taken into account.  The average risk amplification factor  across the 20 banks are found to be  $\rho^{\textrm{EL}}=4.3$, $\rho^{\textrm{VaR}}=4.5$, and $\rho^{\textrm{ES}}=3.2$. Note that the overall levels of commercial credit risk strongly differ across banks --- again potential reasons being differences in business models, etc. as outlined before. Amplification factors also differ across banks.

\begin{figure}[tbh]
	\centering
	\includegraphics[scale=.06, keepaspectratio]{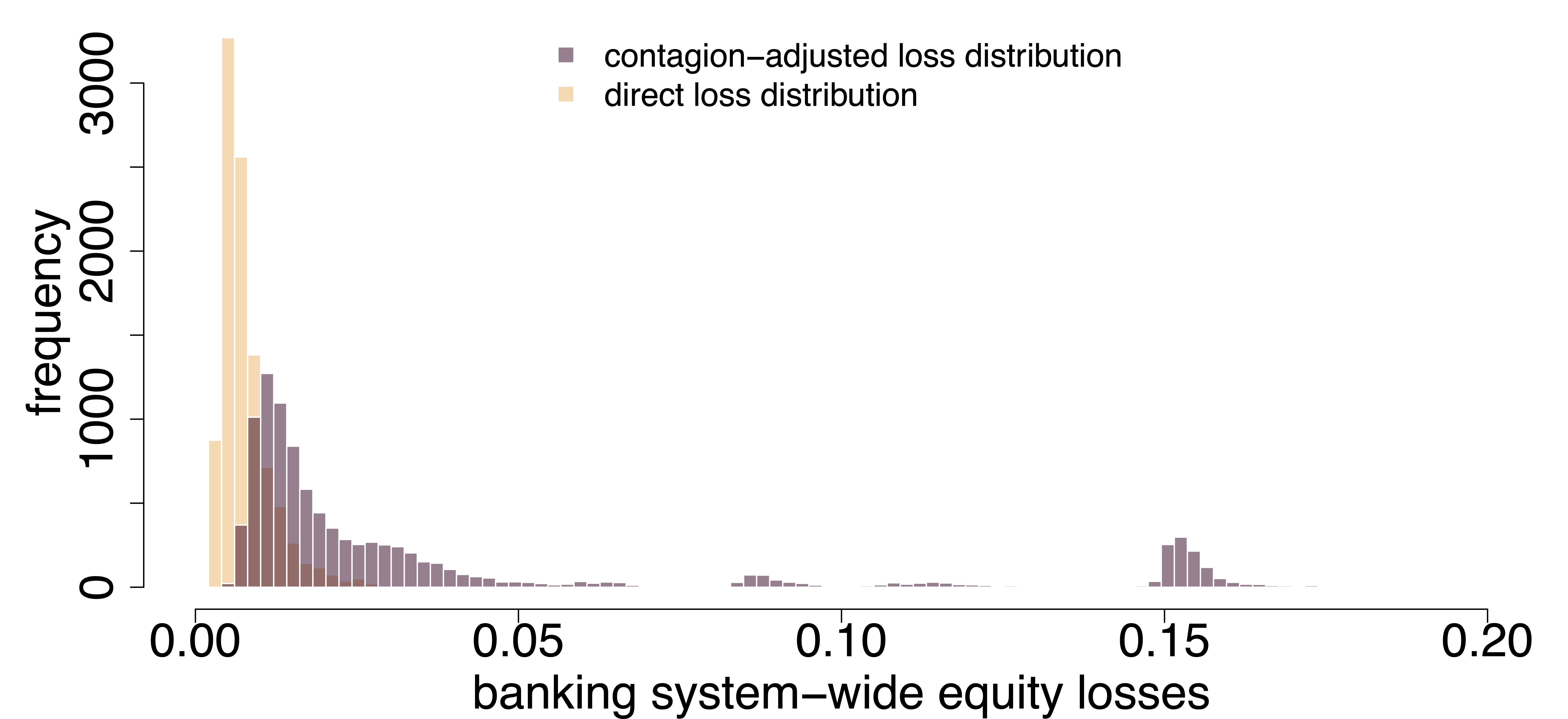}
 	\caption{\textbf{Direct (orange) and contagion-adjusted (brown) loss distributions} across the entire banking layer of $27$ banks. The $x$-axis shows losses to the financial system, relative to the overall equity of all banks, the $y$-axis is the occurrence frequency of losses. The peak around $0.15$ corresponds to shock scenarios in which firms with highest FSRI default. Risk amplification factors for the EL, VaR, and ES are $4.9$, $9.7$, and $7.8$, respectively.
 	}
	\label{lossdistr_two}       
\end{figure}

Finally, in Fig.\ref{lossdistr_two} we present the system-wide histogram  of direct losses, $\mathcal{L}^{\textrm{dir}}(\psi^{\text{M},\ell})$ (orange) and the contagion-adjusted ones, $\mathcal{L}(\psi^{\text{M},\ell})$ (brown), summed over all $27$ banks; see Eq. (\ref{system_loss}). The $x$-axis denotes the fraction of equity that is lost due to an initial shock scenario, $\psi^{\text{M},\ell}$, the y-axis shows the frequency the respective loss. One clearly sees that the two distributions differ substantially. The direct loss distribution is concentrated at low values and has only scenarios where a maximum of around $2.5\%$ is reached. The contagion-adjusted loss distribution is bimodal with most weight centered around $1\%$. It slowly decays with a fat tail and has a second concentration point around $15\%$ of system wide equity losses. In between there are a few scenarios yielding losses of around $9\%$ and $11\%$. This pattern is driven by the shape of the FSRI distribution; see Fig. \ref{FSRI}. The most risky firms cause system-wide losses of around $15\%$ of equity. If one of these firms fails in an initial shock scenario, $\psi^{\text{M},\ell}$, they trigger a supply chain cascade leading to losses of around $15\%$. Similarly, there are a few  high systemic risk firms causing losses of around $11\%$ and $9\%$ of equity, explaining the respective parts in the loss distributions. The risk respective amplification factors for EL, VaR, and ES are $4.9$, $9.7$, and $7.8$, respectively. 

\section{Discussion} \label{discussion}

A thorough assessment of credit risk is central to a sustainable banking sector. Classical credit rating models hitherto do not include information about supply chain networks (SCNs) on the firm level, their structures, and default dynamics in a systematic way. Here we introduced a fully data-driven model (containing a minimum number of free parameters) that allows us to quantitatively estimate how SCN contagion translates to capital losses in the banking sector and to what extent it affects financial stability. The model is a 1:1 agent-based  representation of (almost)every single firm and bank in the national economy of Hungary in 2019, containing more than 240,000 firms, 27 banks, including their balance sheets and income statements, along with more than $1.1$ million supply links and $35,609$ bank-firm loans.  A unique feature of this model is that the 1:1 representation of the real economy is linked to financial system through credit exposures. It allows us to study how risks spread from the real economy (production layer) to the financial layer. We follow the philosophy that parameters in ABMs should not be freely assumed but fully are determined and fixed by the data. This is the reason why we refrain from including additional features to the model that we can not calibrate with available data, such as  details of  the rewiring dynamics. For estimating the production functions of the companies and a simple substitutability dynamics we follow \cite{diem2022quantifying, diem2023estimating}. 
As, at this stage, we do not employ rewiring dynamics the model is not yet predictive or generative. It has been conceived in the spirit of stress testing. It is designed to estimate and monitor the likely costs of hypothetical external shocks, assuming that there are no interventions of governments or central banks. 
It is not designed for medium- and long term predictions but rather for short time scales below the typical times that are needed to restructure the SCN. 

Within this framework our results show that the failure of very few individual firms (those that can cause large cascades of supply chain contagion), can lead to system-wide bank equity losses in the banking sector of around $15\%$. 
These losses can be mostly attributed to supply chain contagion induced defaults of firms (indirect losses); the loan write-offs caused by the initially failing firms (direct losses) only play a minor role. Half of the financial system's losses, approximately $8\%$, caused by an initial failure of one of the riskiest firms, can be linked to firms defaulting only due to their insufficient liquidity buffers. This implies that  increasing liquidity provision to firms that do not face negative equity levels can alleviate already large parts of loan defaults. However, a substantial amount of bank capital losses  (approximately $7\%$) would still be caused by firms with larger losses than equity in response to supply chain disruptions.

Our results also indicate, that those firms causing the largest losses do have no, or relatively small, loans from Hungarian banks (they could either borrow from the capital market or from banks abroad), while firms with the largest loans cause  small indirect supply-contagion-induced financial losses. This implies that firms with the largest bank-loans  might not be the most relevant for monitoring financial stability, but those that can cause substantial indirect defaults. This type of firms would be clearly missed when not taking supply chain contagion into account. It is indicated that regulators could benefit from building up capabilities to monitor supply chain generated and amplified systemic risks. Governments and regulators could collect this type of data as done today already in about a dozen of countries \citep{bachilieri2022firmlevel}. 

Analysis of system-wide bank equity losses shows that the bulk of the indirect losses caused by the systemically riskiest firms are suffered by 10 banks. These make up for around $74\%$ of the entire equity, $85\%$ of loans, but have a substantial number, $93\%$, of corporate clients. This  indicates that  a relatively large number of clients can potentially expose banks to direct and indirect losses, i.e., systemic risk. 

Within our framework we performed a stylized stress test by sampling initial shocks based on randomly failing about $1\%$ of firms (based on their estimated default probabilities) and simulating the spread of the production network contagion to the financial layer. This way we create 10,000 initial shock scenarios and obtain two loss distributions for each bank, one with and the other without SCN contagion. The results suggest a substantial amplification of financial risks, when SCN contagion is taken into account. On average, expected losses of banks are amplified by a factor of $4.3$, value at risk - by a factor of $4.5$, and expected shortfall - by a factor of $3.2$. Amplification factors differ across banks. Nevertheless, if events trigger supply-chain contagion, then direct losses of bank equity are negligible in comparison to the indirect losses caused by SCN contagion. This implies that shock propagation in SCNs can potentially lead to substantially fatter tails in banks' loan defaults, suggesting that capital ratios should factor in these risks.

The observed amplification of credit risk measures by $300-450\%$ unambiguously demonstrates that supply chain contagion is an extremely important factor for a realistic credit risk assessment in a world with increasing numbers of supply chain disruptions. It also is a message that current practice could systematically and grossly underestimates capital requirements and sustainable interest rates when large exogenous shocks trigger SCN contagion. The present study strongly indicates that supply chain contagion poses unexpectedly large systemic risks to financial stability. In practice the default probability estimates (performed by banks) implicitly also account for defaults due to SCN contagion occurring in normal times. This is because the data (past default events in the loan portfolios) contains defaults that originated from SCN contagion, e.g., the failure of a major customer or supplier). On the one hand, our results hint at the fact that SCN contagion actually is a substantial driver of credit risk. On the other hand most data points to  correspond to times where no major SCN contagion has occurred or it has been cushioned by government interventions, such as during the COVID-19 pandemic. Therefore, credit risk might be underestimated. 
Mis-estimations of correlations in housing loans played a major role in causing the financial crises 2007/2008. Ignoring that supply chain networks have the potential to cause substantial default correlations can lead to unexpected financial losses due to neglecting economic mechanisms leading to correlated defaults --- this time of firms.

The presented study has a number of obvious limitations. Foremost, our results depend on a range of assumptions that are relevant for the micro-simulation model. These include the supply-chain contagion mechanism itself, how production disruptions due to supply-chain contagion translate into financial losses for firms, and how the insolvency of firms translates into capital losses for banks. We take a pragmatic approach for firm behaviour. 
In the proposed setting we assume that revenues and material costs are reduced proportional to the production loss that a firm suffers after the SCN contagion. In this way all firms are treated equally without any consideration of their industry affiliation, or nuances in their business models. Given the present data it is not feasible to account for varying transmission mechanisms according to business models. However, future improvements could take into account the  industry sector of firms and other not yet utilized items of their income statements and balance sheets.
A second assumption is the condition under which firms default. We use  firm's equity and short term liquidity (cash, short term claims and securities subtracted by short term liabilities) as buffers against income losses. If the losses exceed one of the buffers we use this as an indicator of default. 
Our current buffer calculation neglects  that firms could, e.g., ask for additional liquidity from banks, or tap additional equity from their owners and, hence, avert a default.
A further assumption is the value of the loss given default (LGD). We assume it to be at $100\%$ of the outstanding principal. This means that we do not take into account the possibility of instalment postponement, restructuring of firms in case of any problems or use of  collateral by banks. However, for the short term view this is not unrealistic as  collection and resolution processes can last years. Further, the production loss of a firm, its bankruptcy and default, and losses of banks happen immediately, without a detailed concept of time being modelled. In this manner our results suggest upper limit of possible losses.
Finally, the data even though unique in coverage has shortcomings. 
The supply-chain network is not entirely complete; small firms and import-export links are missing. We only included the largest banks due to data availability, which, however, should affect results only to a limited extent since the majority of bank-firm loans and loan volume are covered. 

Data driven 1:1 models along the lines presented here can be used as a base to assess governmental recovery policies, such as loan guarantees for firms, revenue loss subsidies that have been implemented, for example during the COVID-19 pandemic. Applying it to specific shock scenarios such as pandemic restrictions or devaluations of assets due to the climate change, would be natural future extensions. Model assumptions would have to be calibrated to the specific initial shock scenario. Maybe the most important step to make the model more practicable, also as a predictive tool, is to include price information that is so-far missing. Further, the contagion between banks needs to be modelled as well as the financing for firms' production activities. Also more detailed behavioural mechanisms for firms and banks should be included in future work.

In a wider context, the demonstrated risk amplification mechanism through the coupling of the financial network to the SCN can be seen as a concrete example that ``networks of networks'' are generally expected to show a significant risk-increase in comparison as defaults on simple networks  \cite{PhysRevLett.107.195701}.

\bibliography{FSRI_article}


\appendix 
\section{Economic Systemic Risk Index} \label{appendixA}

We use the supply-chain network, $W$, ($n\times n$ matrix, $n=243,339$), where $W_{ij}$ is the trade volume between buyer, $j$, and supplier, $i$.  Therefore, total purchase volume of firm $i$, the \textit{in-strength}, is $s_i^{\textrm{in}}=\sum_{j=1}^n W_{ji}$, and total amount of sales, the \textit{out-strength}, is $s_i^{\textrm{out}}=\sum_{j=1}^n W_{ij}$.

Every firm, $i$, is equipped with a generalized Leontief production function, defined as
	\begin{equation} \label{eq_glpf}
		x_i = \min\Bigg[
		\min_{k \in \mathcal{I}_i^\text{es}} \Big[ \frac{1}{\alpha_{ik}}\Pi_{ik}\Big], \: 
		\beta_{i} + \frac{1}{\alpha_i} \sum_{k \in  \mathcal{I}_i^\text{ne}} \Pi_{ik} ,  \; \frac{1}{\alpha_{l_i}}l_i, \; \frac{1}{\alpha_{c_i}}c_i \;   \Bigg] \, .
	\end{equation}
$\mathcal{I}_i^\text{es}$ is the set of essential inputs and  $\mathcal{I}_i^\text{ne}$ is the set of non-essential inputs of firm $i$. 
The parameters $\alpha_{ik}$ are technologically determined coefficients,  $\beta_{i}$ is the production level possible without non-essential inputs $k \in \mathcal{I}_i^\text{ne}$ and  $\alpha_{i}$ is chosen to interpolate between the full production level (with all inputs) and $\beta_{i}$. All parameters are determined by $W$, $\mathcal{I}_i^\text{es}$ and $\mathcal{I}_i^\text{ne}$. Note, that labour, $l_i$ and capital, $c_i$, are not explicitly modelled, due to the short-term perspective of our model. The initial labour shock is modelled as the remaining production level, $\psi_i$. 
We simulate how the initial supply and demand reductions spread downstream to the direct and indirect buyers and upstream to the direct and indirect suppliers of the initially affected firm, by recursively updating the production levels of all firms in the network. 
We update for each firm $i$ the production output at $t+1$, given production levels of its suppliers, $h^\text{d}_j(t)$, at time $t$ as
\begin{eqnarray} \label{eq_downstream}
	x_i^{\text{d}}(t+1) & = &   \min\Bigg[
	\min_{k \in \mathcal{I}_i^\text{es}} \left( \frac{1}{\alpha_{ik}}\sum_{j=1}^{n} W_{ji} h_j^\text{d}(t) \delta_{p_j,k}\right), \\
	  & &  \beta_{i} + \frac{1}{\alpha_i} \sum_{k \in  \mathcal{I}_i^\text{ne}}  \sum_{j=1}^{n}   W_{ji} h_j^\text{d}(t) \delta_{p_j,k}, \, \psi_i x_i^\text{}(0)
	\Bigg] \quad .  \notag
\end{eqnarray}
The production output of firm $i$ at $t+1$, given the production level of its customers, $h^\text{u}_l(t)$, at time $t$ is computed as
\begin{equation}\label{eq_upstream}
	x_i^{\text{u}}(t+1) = \min\Bigg[ \sum_{l=1}^{n} W_{il}h_l^\text{u}(t) , \, \psi_i x_i^\text{}(0) \Bigg] \quad. 
\end{equation}
For details and the full algorithm we employ see the methods section in \cite{diem2022quantifying}. 

The algorithm can be used to asses a systemic importance of every firm in the SCN. By defaulting a firm, i.e., using a scenario $\psi^{S,j}$, and running the supply chain contagion model, we end up with remaining production levels at time $T$ of firms in the network represented by a vector, $h(T)\in \left[0,1\right]^n$. The \textit{economic systemic risk index (ESRI)} of a firm $j$ is now defined as
\begin{equation}
    \textrm{ESRI}_j \equiv \sum_{i=1}^n\frac{s_i^{\textrm{out}}}{\sum_{l=1}^n s_i^{\textrm{out}}}(1-h(\psi^{S,j},T)) \quad .
    \label{esri}
\end{equation}

The losses in response to more general shock scenarios, $\psi^{M}$, are modelled with the same shock propagation dynamics as is demonstrated in \cite{diem2023estimating} for COVID-19 shock scenarios.

\section{Central bank methodology to PD-estimation for all Hungarian firms liable to corporate tax} \label{appendixB}

The default probability (PD) estimation originates from the methodology developed by \cite{banai2016magyar} and refined by \cite{burger2022defaulting}. This methodology estimated PD values for SMEs included in the annual Corporate Income Tax database of the Hungarian National Tax and Customs Administration (‘NAV Társasági Adó’), and had at least one loan or leasing contract (i.e. they were registered in the Credit Registry (KHR) for corporates). The model was trained using observations from the years between 2007 and 2019. For the purpose of this current paper, the original estimation procedure was simplified. While the model was still trained on companies with loan contracts, the goal of the simplification was to make PD-predictions possible for all Hungarian corporates, not just the ones with loan contracts. Hence, all loan-contract level information was removed from the estimator function.

The original methodology of Banai et al (2016) considered a company to be in default if its instalment payment was at least 90 days overdue. Burger (2022) refined this step by considering defaults only if at least 99 percent of all loan amounts was in default. One year PD was predicted, which means that the probability of default in the four quarters following an observation date was estimated. In order to harmonise this approach with the data, annual values of the Corporate Tax database were converted to quarterly values. Balance sheet values were linearly interpolated, while income statement positions were divided by four. In the absence of further information on seasonality and similar metrics, this was considered as an acceptable compromise between modelling needs and data availability. All these transformations boiled down to 163 thousand tax-liable, credit-taker, distinct SMEs from 2007 till 2019, represented by 3.4 million quarterly observations.

{The PD-estimator relied on a set of macroeconomic variables (3-months BUBOR, quarterly changes to the HUF-EUR exchange rate, unemployment levels), loan contract-level variables aggregated to company level (such as foreign currency-denominated loan flag, longest loan maturity, sum of all loan values in HUF) and company-level financial variables.} To keep the independent variables with explanatory power only, the LASSO-dimension reduction algorithm was applied. And while the final estimator function was a logistic regression, non-linear relations were minimal, as an analysis with the help of gradient boosting showd.

This present analysis required alterations to the methodology described above. First, all industries but financial corporations were kept in the data, which were originally excluded. Second, the loan contract-related explanatory variables of the original PD-estimator were removed, since the aim of the model was to make predictions possible for all corporates, not just for those with loans. Finally, the definition of active companies was expanded: now companies with zero revenues but with positive total assets were kept in the training set. This training data was used for PD-estimation that was prepared with the help of LASSO and a simple logistic regression.

Using the model trained in this way, PDs were estimated for about 143,000 companies liable for Corporate Tax, with or without a loan contract. There used to be, however, a sizeable mass of companies using alternative taxation formats (EVA, KATA, KIVA), in our dataset about 52,000, before the recent overhaul of the corporate taxation system, hence the algorithm described did not generate PDs for them. To overcome this, a separate, simple PD-estimator was developed, on the training data again. It was based on an elementary, but available information, such as industrial affiliation, company age, revenue category, a category based on the number of employees, and finally, legal form. While this simplification is certainly not the most sophisticated, it provides reasonable estimations of PD values. Finally, PDs for all 195,000 companies were generated.

\section{Correlation of FSRI with sales and loans} \label{appendixC}

\begin{figure}[tbh]
	\centering
	\includegraphics[scale=.06, keepaspectratio]{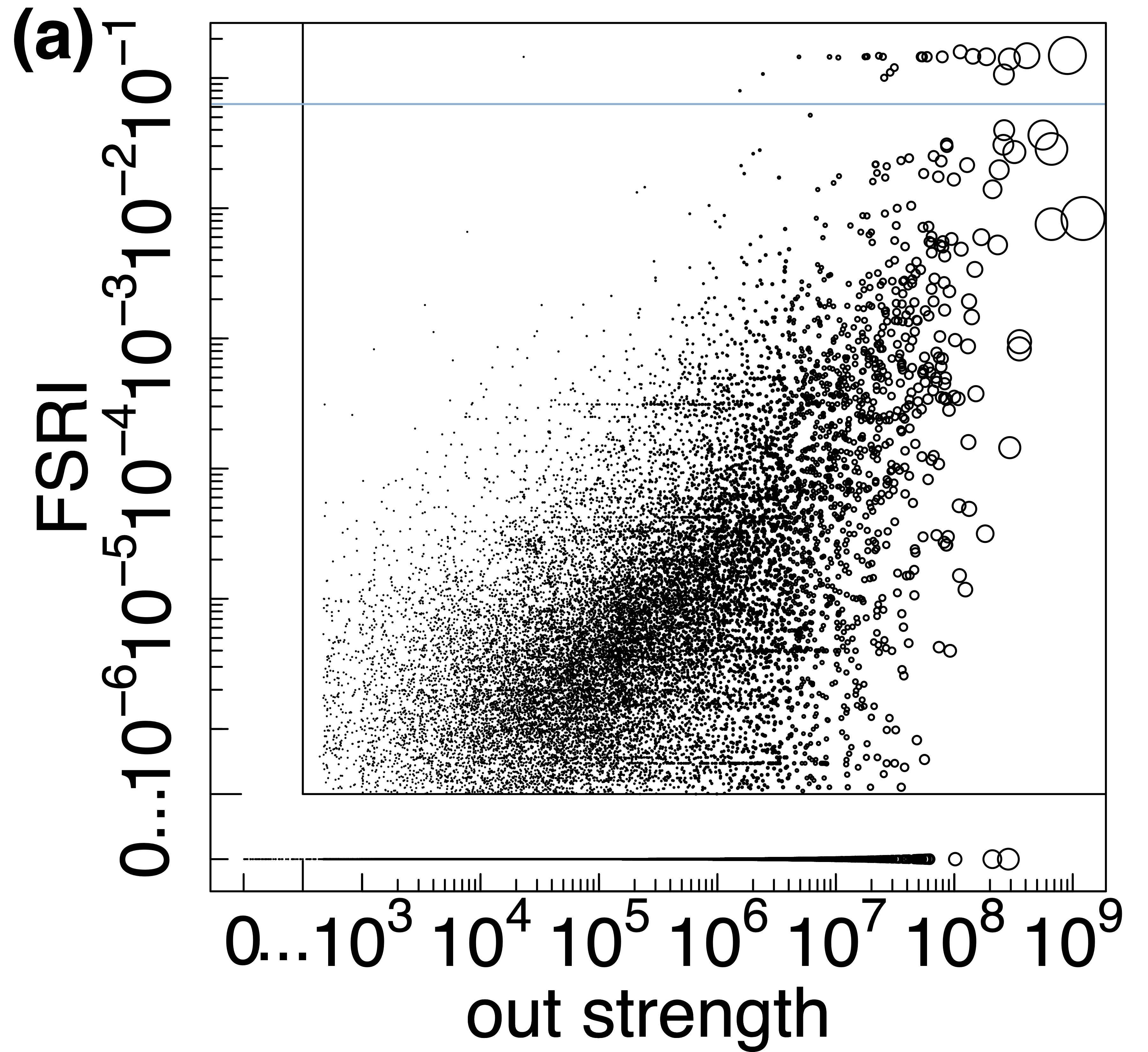}
	\includegraphics[scale=.06, keepaspectratio]{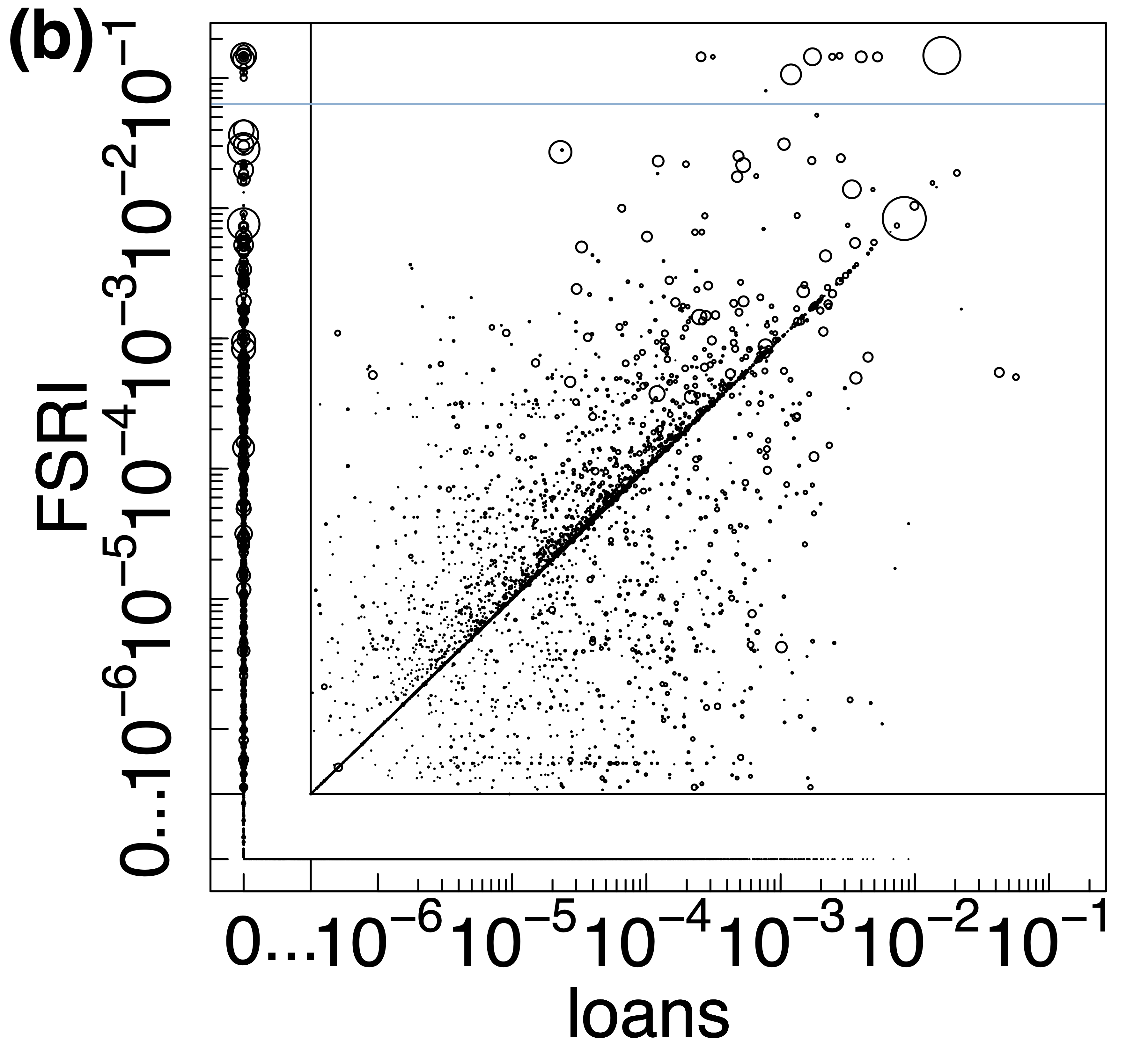}
 	\caption{\textbf{FSRI against total sales and loans of firms.} The first log-log (plus zero section) scatter plot \textbf{(a)} shows the out strength of firms ($x$-axis) against their FSRI values ($y$-axis). Bubble size is proportional to the out strength (total sales) of a company. The horizontal (blue) line is plotted at $0.06$ indicating the most risky firms which are above it. \textbf{(b)} shows loans of firms divided by the total equity of $27$ banks ($x$-axis) against FSRI ($y$-axis). Bubble size corresponds to out strength. Values on the diagonal are equivalent to the direct losses (orange bubbles) plotted in Fig.\ref{FSRIdirectindirect}(b).
  }
	\label{FSRIoutstrloans}       
\end{figure}

\cite{diem2022quantifying} show that economic systemic risk of a firm is not necessarily proportional to total sales. In Fig.\ref{FSRIoutstrloans}(a), we see that this holds also for FSRI. Firms with very small out strength (above the blue line) can carry a substantial systemic risk. Fig.\ref{FSRIoutstrloans}(b) shows FSRI against loans of firms on a log-log scale. As we mention in Section \ref{data}, majority of firms in the SCN don't have loans. Nevertheless, they can cause losses to banks indirectly indicated by the column of bubbles on the left hand side of the plot. The firms on the top of the column, upper left corner, are firms with high systemic risk and with no loans in the country. Other firms above the blue line are also the most risky firms from the plateau in \ref{FSRI}. Companies that are on the diagonal cause only direct losses equal to their own loans. All other firms, above and under the diagonal, cause indirect losses as well. Altogether, there is no clear correlation of loan size and FSRI.

\section{Contagion-adjusted probability of default} \label{appendixD}

\begin{figure}[tbh]
	\centering
	\includegraphics[scale=.06, keepaspectratio]{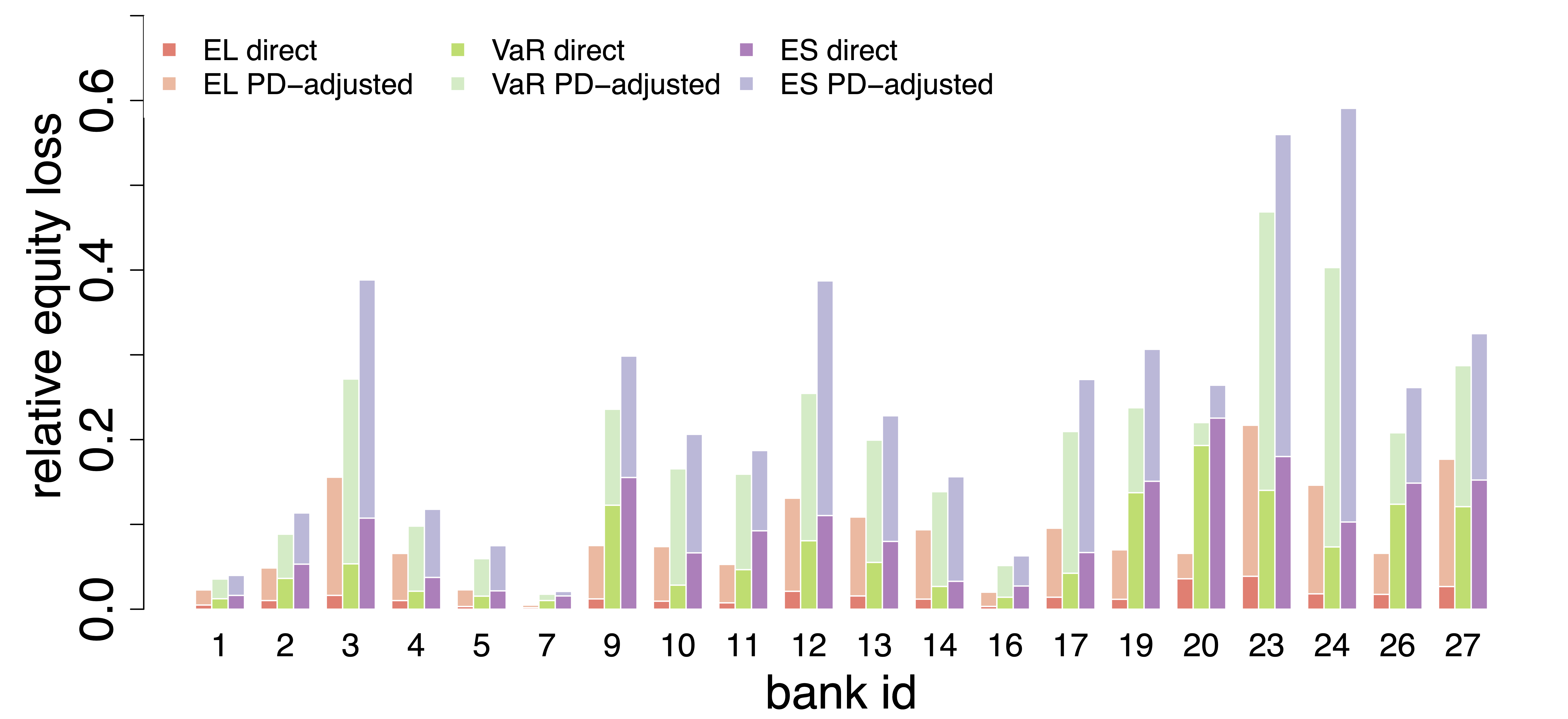}
 	\caption{\textbf{Expected loss, value at risk, and expected shortfall of banks from direct and PD-adjusted loss distributions.} As  Fig.\ref{barplot_MC}, on $x$-axis represents bank indices and $y$-axis shows risks relative to equity of each bank. We skip banks n.6, 8, 15, 18, 21, 22, 25 which have less than $30$ clients. Risk measures EL, VaR, and ES are denoted by red, green and purple colours respectively, for direct and PD-adjusted loss distributions in dark and light shades respectively. The bar-plot is not stacked, meaning that dark bars are always plotted in front of light bars. Average risk amplification factors are $\rho^{\textrm{EL}}=6.1$, $\rho^{\textrm{VaR}}=3.4$, and $\rho^{\textrm{ES}}=2.8$.
  }
	\label{barplot_PD}       
\end{figure}

\begin{figure}[tb]
 \centering
    \includegraphics[scale=.04, keepaspectratio]{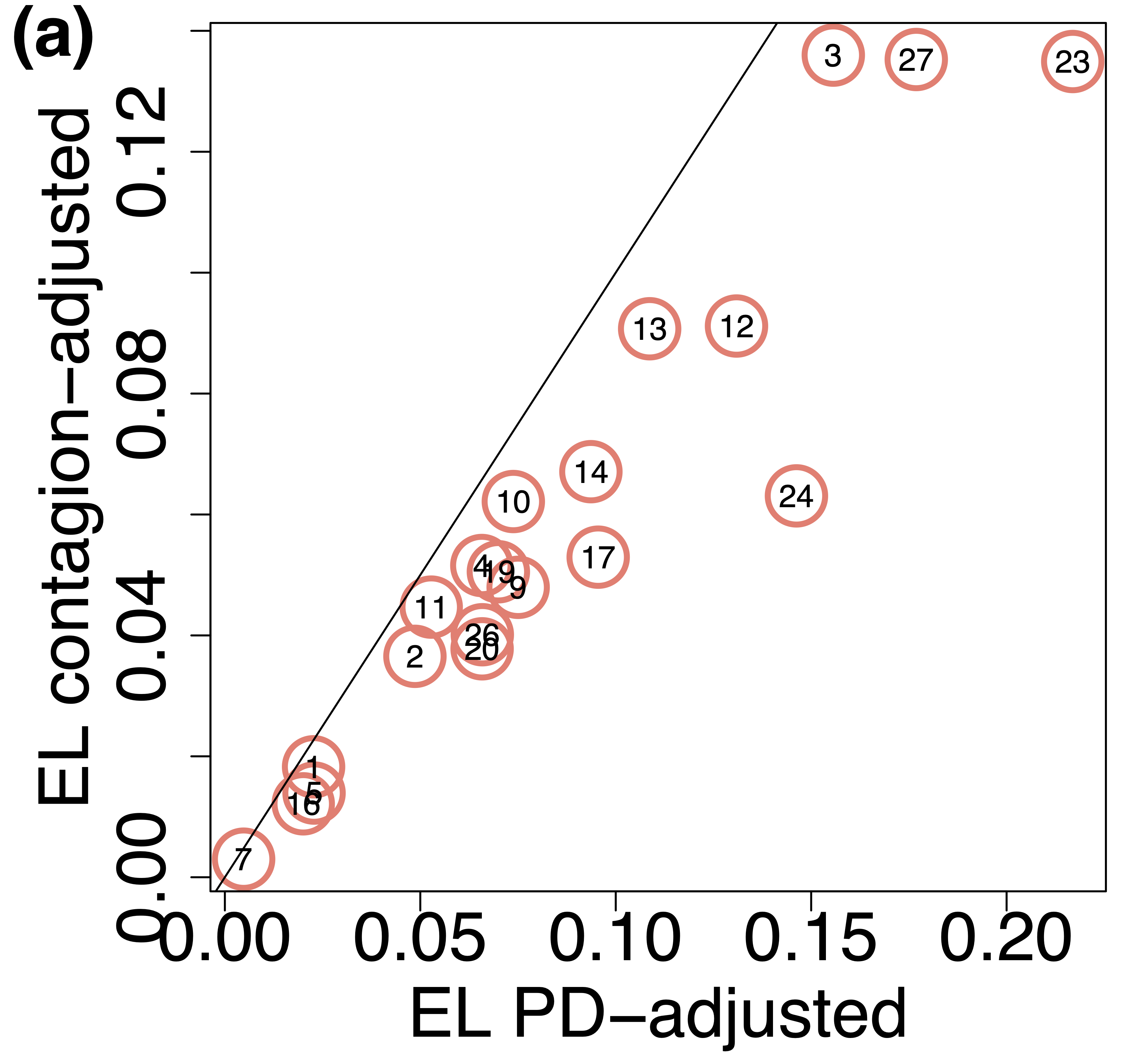}
    \includegraphics[scale=.04, keepaspectratio]{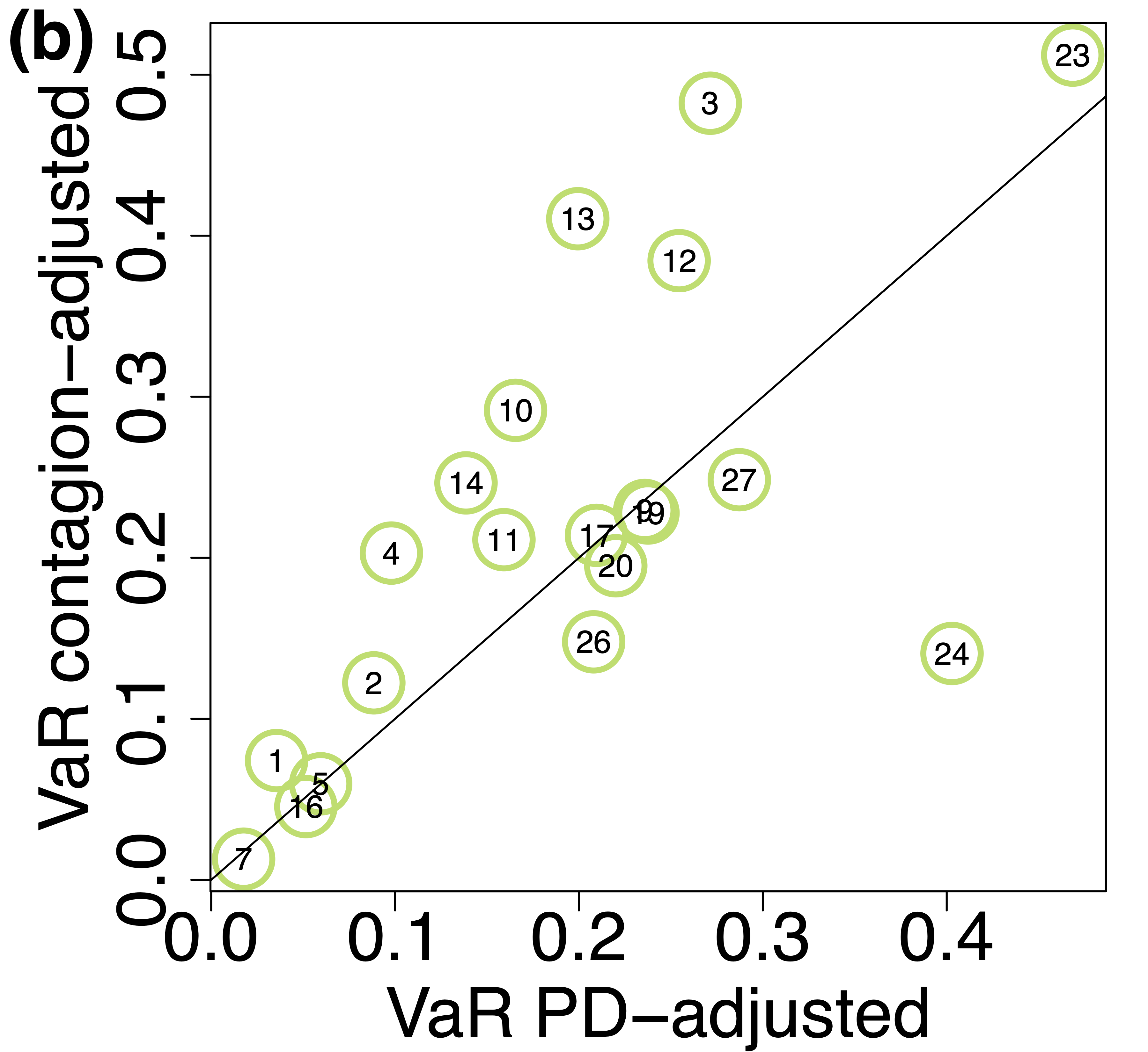}
    \includegraphics[scale=.04, keepaspectratio]{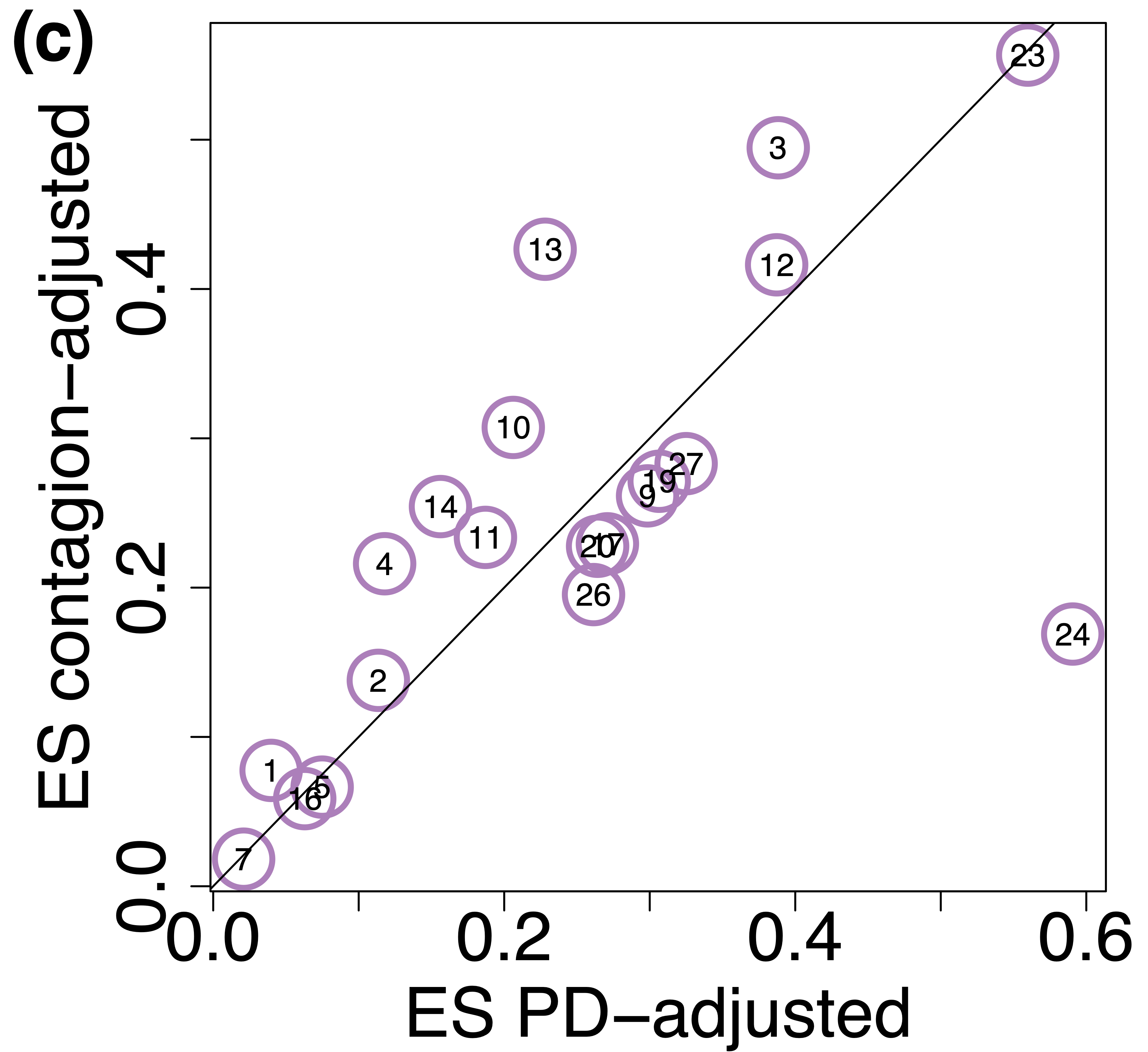}
 \caption{\textbf{Comparison of two methods for the supply chain contagion-stressed loss distributions.} The three panels the show correlation of the results presented in Fig.\ref{barplot_MC} (contagion-adjusted method) and Fig.\ref{barplot_PD} (PD-adjusted method), for EL, VaR, and ES in \textbf{(a)}, \textbf{(b)}, and \textbf{(c)}, respectively.  } 
 \label{comparisonPlot}
\end{figure}

\begin{figure}[tb]
	\centering
	\includegraphics[scale=.06, keepaspectratio]{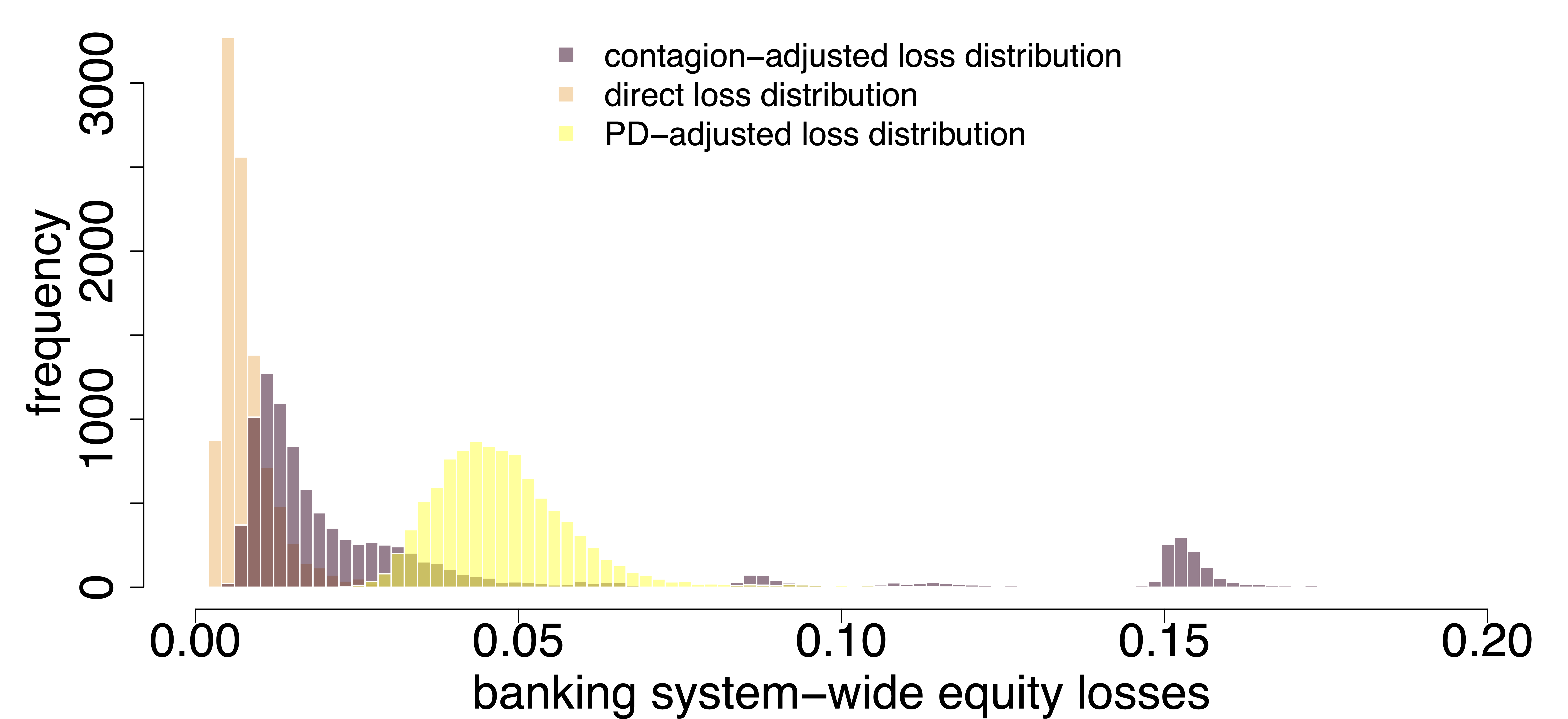}
 	\caption{\textbf{Direct, contagion-adjusted and PD-adjusted loss distributions}. Distribution of direct losses (orange), contagion-adjusted losses (brown), and PD-adjusted losses across the whole banking layer of $27$ banks. The $x$-axis show losses of financial system relative to the overall equity of all banks, the $y$-axis is the frequency of losses. The peak around $0.15$ corresponds to shock scenarios where firms with the highest FSRI default. Risk amplification factors from Eq. (\ref{riskampl}) for EL, VaR, and ES of PD-adjusted and direct loss distributions are equal to $6.2$, $4.2$, and $3.9$ respectively. Risk amplification factors for EL, VaR, and ES of contagion-adjusted and direct loss distributions are equal to $4.9$, $9.7$, and $7.8$ respectively.
 	}
	\label{lossdistr_three}       
\end{figure}

Here we propose an alternative method for assessing contagion-adjusted system wide and bank specific loss distributions. Based on the supply-chain contagion after the initial failure of each single firm --the calculations underlying FSRI-- we identify the so-called \textit{critical suppliers and buyers} of every firm, $i$. These are direct and indirect suppliers and buyers, which in case of their failure would cause a default of firm $i$. With this information, we can adjust the ``idiosyncratic'' probability of default of firm $i$, $\text{pd}_i$ by taking the ``idiosyncratic'' probabilities of default of its critical suppliers and buyers into account. This gives us the the \textit{contagion-adjusted probability of default}. $q_i$ ($q_i \geq \text{pd}_i $). Based on these probabilities, $q_i$, we sample $10,000$ Bernoulli random vectors of size $243,339$ firms with the success probability, $q_i$, indicating that firm $i$ failed. Note that before we simulated 
%
$\psi_i^{\text{M},\ell} \sim {Ber}(\text{pd}_i)$ 
and used this as initial shock of the supply chain contagion algorithm. In this approach the supply-chain contagion channel is implicitly considered by the adjusted PDs, $q$. We calculate again the losses $\mathcal{L}_k$ and $\mathcal{L}$ to generate loss distributions of every bank separately and of the whole financial sector, see Fig.\ref{barplot_PD} and Fig.\ref{lossdistr_three}. In Fig.\ref{comparisonPlot} we present comparison of results from two methods. Risk amplification factors (\ref{riskampl}) in the case of PD-adjusted method for EL, VaR and ES are 6.2, 4.2, and 3.9 respectively. We see, that the second method unnecessarily captures contagion risk in expected losses. The contagion-adjusted loss distribution given by the first method also increases expected losses, but risks given by the systemically important firms are captured by tails as a less probable events, which they indeed are. 

Our aim is to incorporate the supply chain (SC) contagion channels leading to defaults, as described in the Section \ref{method}, into conventional credit risk measure - the probability of default (PD). Note, that information about SC links is not involved in the PD model of the central bank, described in the \ref{appendixB}. Nevertheless, we do not intend to develop any new model which would include supply chain links as additional variables, and rather adjust the existing values of PDs.  We use a simple observation arising from our method \ref{method}, that failures of some firms lead to defaults of another ones via SC contagion channels. Therefore, PD of the defaulted firm should increase proportionally to PD of the failed firm. We assume, that probabilities of failures of firms (i.e. events that start contagion) are equal to their respective PDs estimated by the central bank model.

Let us recall some notation from the main part. In Section \ref{method} we discussed the production network contagion initiated by the failure of a single firm, $j$, represented by an initial shock, $\psi^{\text{S},j}$. Contagion leads to production losses of some firms in the network and, in some cases, to bankruptcy and default. The binary vector, $\chi=\chi(\psi^{\text{S},j})$ (\ref{chi}), indicates defaulted firms, $\chi_i(\psi^{\text{S},j})=1$ means that a failure of a firm $j$ leads to a default of a firm $i$. If the latter is true, firm $j$ will be called a \textit{critical supplier or buyer} of the firm $i$. Events where a critical supplier or buyer of a firm defaults are equivalent to events where the firm itself defaults. 

Thus, the \textit{contagion-adjusted probability of default}, $q_i$, of the firm $i$ with  $\eta=\eta(i)$ critical links is a joint probability over all events where at least one of the critical buyers and suppliers, or the firm itself, defaults. Let $p_i \equiv \textrm{pd}_i$ ($i=1,2,...,\eta(i)+1$) denote default probabilities of critical suppliers or buyers of the firm $i$ including the firm itself, then
\begin{equation}
    q_i = \sum_{k=1}^{2^{\eta+1}-1} \prod_{l\in I_k} p_l \prod_{j \in I\setminus I_k} (1-p_j) \quad ,
    \label{PD_network}
\end{equation}
where $I$ is a an index set with $\eta+1$ elements, and $I_k \subseteq \mathcal{P}(I)\setminus \emptyset$. $\mathcal{P}(I)$ is the power the index set, with $2^{\eta+1}$ elements. Indeed, in Eq. (\ref{PD_network}) the sum is over all possible combinations of default and non-default of critical nodes, except for the situation when no company defaults.

For example, if a firm $i\equiv a$ has one critical buyer $j \equiv b$ and the independent default probabilities are estimated by the central bank as $p_a$ and $p_b$, then $\mathcal{P}(I)\equiv\{\{\emptyset\},\{a\},\{b\},\{a,b\}\}$ is the power set with $2^2$ elements, $I\equiv\{a,b\}$ and $I_k \subseteq \{\{a\},\{b\},\{a,b\}\} $ with $k=1,2,3$.
The probability of a default $q_a$ of the firm $a$ conditional on the critical network links can be obtained from Eq. (\ref{PD_network}) as
\begin{equation}
    q_a = p_a p_b + p_a(1-p_b) + (1-p_a)p_b \quad .
\end{equation}
However, we can quantify $q_a$ in a more convenient way. Summation in Eq. (\ref{PD_network}) over $2^{\eta+1}$ elements, including the omitted term,
\begin{equation}
    \prod_{j=1}^{\eta+1} (1-p_j) \quad , 
\end{equation}
when no firm defaults, is equal to $1$. It is because $2^{\eta+1}$ elements span the complete probability space. Therefore, we can rewrite Eq. (\ref{PD_network}) as  
\begin{equation}
    q_i = 1 - \prod_{j=1}^{\eta+1} (1-p_j) \quad ,
    \label{PD_net}
\end{equation}
where $q_i \geq p_i$ for every firm $i$ from the supply chain network. The probability of default $q_a$ from the example is thus
\begin{equation}
    q_a = 1 - (1-p_a)(1-p_b) \quad .
\end{equation}

Finally, we can use the \textit{contagion-adjusted default probabilities}, $q_i\textrm{ }(i\in n)$, to generate  \textit{PD-adjusted loss distributions} of banks and compare them with the contagion-adjusted loss distributions in Fig.\ref{barplot_MC}-\ref{lossdistr_two}. Again, a Monte Carlo simulation of random defaults sampled from to $q_i$s is performed. In every trial (out of $10,000$) losses of a bank are obtained as a sum of all loans of its defaulted clients (i.e. the \textit{loss given default}, LGD$=100\%$). EL, VaR, and ES (light red, light green and light purple respectively) of PD-adjusted loss distributions in comparison with the same measures of direct loss distributions (Fig.\ref{barplot_MC}) are presented in Fig.\ref{barplot_PD}. The average risk amplification factors are $\rho^{\textrm{EL}}=6.1$, $\rho^{\textrm{VaR}}=3.4$, and $\rho^{\textrm{ES}}=2.8$. 

If we compare these average amplification factors with amplification factors from contagion-adjusted loss distributions presented in Fig.\ref{barplot_MC}, we see, that the $\rho^{\textrm{EL}}$ is higher for the PD-adjustment method, while the $\rho^{\textrm{VaR}}$ and $\rho^{\textrm{ES}}$ are higher in case of the contagion-adjusted approach. Indeed, from the Fig.\ref{comparisonPlot}(a) it is clear that ELs from PD-adjusted method are higher or equal for all banks. In the case of VaR and ES, the pattern in Fig.\ref{comparisonPlot}(b)-(c) does not change fully, nevertheless, risks are higher from the contagion-adjusted method for the majority of banks. This means that the contagion-adjusted method can capture SC shocks better, which are less probable, in the tails of loss distributions. Figure \ref{lossdistr_three} is an analogue of Fig.\ref{lossdistr_two} with the added PD-adjusted loss distribution (yellow) of the
whole banking layer. The x-axis show losses of financial system relative to the overall equity of
all banks. Risk amplification factors from Eq. (\ref{riskampl}) for EL, VaR, and ES of PD-adjusted and direct loss distributions are equal to $6.2$, $4.2$, and $3.9$ respectively.

\end{document}